# Increasing accessibility by public transport benefits local economy: the effect of a new metro line in Rome

Francesco Marzolla[1,2,3,*], Bruno Campanelli[1,2,3], Hygor Piaget Monteiro Melo[1,2,4,5], Matteo Bruno[1,2], and Vittorio Loreto[1,2,3,6]

[1]Sony Computer Science Laboratories - Rome, Joint Initiative CREF-SONY, Centro Ricerche Enrico Fermi, Via Panisperna 89/A, 00184, Rome, Italy
[2]Centro Ricerche Enrico Fermi (CREF), Via Panisperna 89/A, 00184, Rome, Italy
[3]Sapienza Univ. of Rome, Physics Dept, Piazzale A. Moro, 2, 00185, Rome, Italy
[4]Postgraduate Program in Applied Informatics, University of Fortaleza, 60811-905, Fortaleza, CE, Brazil
[5]Núcleo de Ciência de Dados e Inteligência Artificial (NCDIA), University of Fortaleza, 60811-905, Fortaleza, CE, Brazil,
[6]Complexity Science Hub, Josefstädter Strasse 39, A 1080, Vienna, Austria
[*]Corresponding author: `francesco.marzolla@uniroma1.it`

October 3, 2025

**Abstract**

This study investigates the economic impact of Metro C, a major expansion of Rome's metro system. Using a difference-in-differences (DID) approach within a multiplicative framework, the research quantifies how increased accessibility influenced local economic activities. The results show a statistically significant rise in the number of economic activities in areas affected by the new line. A mild decline in economic diversity suggests the emergence of spatial clustering of similar activities. A dedicated analysis of microenterprises, which represent the majority of the dataset, examines changes in employment and GDP associated with the new infrastructure. The observed zone-level correlation between accessibility gains and growth in economic activities also offers a basis for generalizing the findings beyond the specific case of Metro C. Overall, the case study shows that public transport investments aimed at boosting sustainable mobility can also generate positive spillover effects on the local economic fabric.

## Introduction

The transport sector accounted for approximately 22% of the global $CO_2$ emissions in 2023 [1], and, according to the International Energy Agency (IEA), it is the economic sector that must undergo the most significant behavioural changes in the near future to achieve carbon neutrality by 2050 [2]. In urban areas, these changes involve shifting from car usage to public transport, alongside cycling walking and ridesharing. Within this shift, urban rail systems play a critical role. As the most energy-efficient and, at present time, least carbon-intensive mode of mass transport [2], rail systems are central to decarbonizing mobility. Metro ridership has been steadily increasing worldwide (excluding the COVID-19 years) [3], yet, to meet climate goals, the share of passenger rail must nearly double (to 20%) by 2050 [2], with urban rail expected to drive much of this growth. Cities could therefore improve their metro lines connections in the near future, and this study aims at characterizing the socioeconomic effect of such a change.

Various studies have explored the relationship between transport infrastructure and local economic development, particularly in urban settings. The prevailing consensus is that new metro lines tend to generate positive externalities by improving accessibility and fostering economies of agglomeration [4]. The latter can be defined as "the decline in average cost as more production occurs within a specified geographical area" [5]. Reduced travel times improve access to jobs and expand the spatial scale over which firms and households can efficiently learn, match, and share resources. Indeed, empirical studies show that reducing travel times deliver gains in productivity [6, 7]. In other words, transport investments enlarge the geographical area within which firms are effectively connected, through reductions in travel times or travel costs [7]. Agglomeration is therefore inherently a function of accessibility [4] and it is likely to generate positive economic gains [7]. Areas with improved access were found to experience increases in the number of firms, coupled with sectoral specialization [8, 9].

Building on this body of work, our case study investigates whether the economic benefits of improved accessibility through public transport investments can also be found in Rome. Specifically, we evaluate the impact of the recent construction of Metro C on local economic activities. This metro line opened in stages between 2014 and 2015 and was connected to the rest of the city's metro system in 2018. It now connects the eastern periphery of the city with the centre. A schematic map of it can be seen in Fig. 1, where Metro C stations are marked with blue dots.

Rome's metro infrastructure is limited, with only 1.4 km of metro lines per 100 000 residents, significantly less than other European capitals such as Madrid, Berlin, London, and Milan [10]. This shortage contributes to high car dependency (63 cars per 100 residents) and severe congestion, with average commuting time three times longer than the average Italian commuting time [11]. The addition of metro line C expanded Rome's metro system by 44% in length, a step toward improved sustainability, though further development is still needed to achieve sustainable mobility [12].

Previous research provides several examples in which new transit infrastructure has influenced local economic activities by changing levels of accessibility. For example, a study on the construction of the Metrosur line in the suburbs of Madrid [13] found that economic activities increased over time in areas located near the new metro stations. The study also showed the presence of spatial clusters, with retail activities displaying the highest pattern of concentration. A case study conducted in London [14] found that the extension of a metro line had a generally positive direct impact on both the productivity and the number of firms located near the newly established stations. Similarly, a recent study on Urumqi [15] reported a positive effect of a newly constructed metro line on the concentration of commercial facilities around its stations. An equally recent study [16] found that the construction of a light rail line in Florence positively affected retail density along the street where it was built and in surrounding streets, especially for stores selling non-durable goods. Another investigation [17], surveyed the opinions of Thessaloniki citizens about a new metro station opening in the city. The vast majority (85%) of respondents believed that local businesses would see an increase in customers once the metro line became operational, and many considered it likely that they would visit the area more frequently.

Here, we conduct a similar investigation for Rome, assessing whether these positive economic patterns, expected in Thessaloniki and found in Madrid, London, Florence and Urumqi, also occurred in the Italian capital. We begin by measuring the accessibility increase due to the introduction of the new metro line. We then select the neighbourhoods impacted by the infrastructure as the ones whose accessibility rose the most. We then assess the impact of Metro C on economic activities through a difference-in-differences (DID) approach, i.e. comparing the impacted neighbourhoods ("treated group") before and after the opening of the new metro line to selected not impacted neighbourhoods ("control group"). Usually the DID estimator, quantifying the causal effect of a treatment, is derived using a linear model [18], here instead we are using a model implying a multiplicative effect of the treatment on the outcome variable. We find a statistically relevant increase in the number of economic activities in treated areas, together with hints of clustering. We finally evaluate the extra GDP and jobs created by the micro-enterprises which have been attracted by the new infrastructure.

# Results

### Accessibility increase due to the introduction of Metro C

First, we evaluate how the introduction of Metro C affects accessibility for economic activities. Specifically, we measure the change in the number of people who can easily reach a given point in Rome via public transport—before and after the introduction of the new metro line—using a public transport efficiency metric [19], whose calculation is explained in detail in Methods. In Fig. 1a, this variation in accessibility is shown for each urban zone, with colours encoding the magnitude of change.

### DID estimation of the impact of Metro C on density of economic activities

To evaluate the impact of Metro C on economic activities we use a difference-in-differences (DID) approach [20]. To do so we need to identify a group of urban zones that receive the treatment whose effect we want to evaluate, in this case, the introduction of Metro C, and a group of urban zones that do not, but experience some or all of the other influences that affect the first group. The former group is called "treated" and the latter "control". We classify an urban zone as treated if, after the construction of Metro C, it could be easily accessed by a significant additional number of people. Treated zones are coloured in (transparent) yellow in Fig. 1b.

Urban zones have been divided in five equal-sized classes, on the basis of their density of economic activities in 2013, one year before the first openings of the metro line. We will refer in the following to these classes with numbers from zero to four, where class zero is the least dense quintile. In Fig. 1b, different shades of blue define the five classes. Metro stations that would open in the following years are marked by blue dots, while grey ones represent stations already present in 2013. The urban zone of Torrespaccata is coloured in black because data on economic activities there contain an outlier and therefore this zone has been discarded from the study.

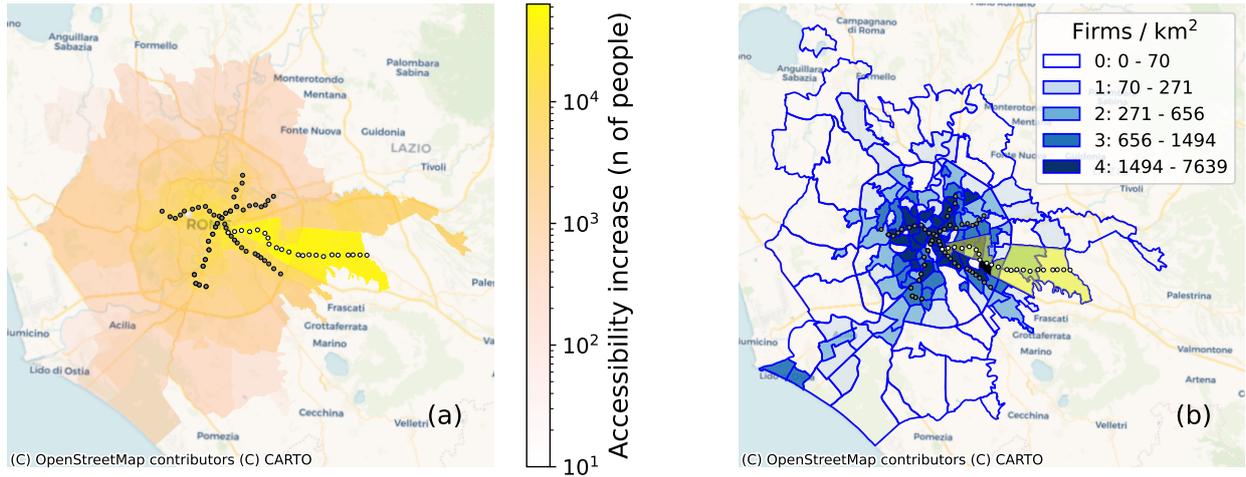

Figure 1: **Accessibility increase of urban areas and selected treated group.** In panel (a), the effect of the introduction of Metro C is measured as the average variation of local accessibility (the number of people who can reach a point using public transport during a typical day) for each zone. Panel (b) shows the density of economic activities and, highlighted, the treated zones. We group the urban zones of Rome in 5 classes based on their density of economic activities in 2013. Classes are marked by shades of blue, Metro C stations are marked with blue dots and treated urban zones are highlighted in yellow. Zones excluded from the study are filled in black.

Inside each class, we assess the impact of Metro C by comparing the variation in the number of economic activities in treated zones with of the average variation in the number of economic activities in control zones as baseline, rescaled. In Fig. 2a we show the average percentage variation in the number of economic activities in treated zones relative to the control group zones. This variation, in the DID framework, estimates the effect of Metro C on economic activities. See Supplementary Information (SI) for the trends of the absolute values of economic activities of the different classes, each with its respective baseline, and for the estimated numbers of economic activities attracted by Metro C for each treated urban zone and for each year. Treated urban zones show an increased economic growth, regardless of the class they belong to. Nonetheless, zones with the smallest density of pre-existing economic activities (class 1) among those connected by the new metro seem to be less affected by the enhancement of the economy due to the new infrastructure. On average, treated zones experienced a $(5 \pm 4)\%$ increase in the number of economic activities by 2018 due to the introduction of Metro C, although there are substantial fluctuations in trends across different zones.

The global number of economic activities attracted because of the opening of Metro C is shown in Fig. 2b. The plot shows for each year the sum over all treated zones of the difference in the number of economic activities between the zone and the appropriate baseline. Shadowed regions span two standard deviations. A $\chi^{(2)}$ test confirms a positive influence of Metro C on the number of economic activities near metro stations, with a p-value less than 0.01.

In Fig. 2c, we cross the DID estimation of the effect of Metro C on the number of economic activities with the increase in accessibility brought by the infrastructure for each urban zone of Rome. On the horizontal axis, we show the accessibility increase due to Metro C, encoded by colour in Fig. 1a, while on the vertical axis, the difference between the number of economic activities present in 2018 in each urban zone and its appropriate baseline. The blue line represents a kernel non-parametric regression and the shadowed region corresponds to a 95% Confidence Interval (CI). The regression, and therefore the firms number variation, is not statistically different from zero for the vast majority of the urban zones of Rome, having an insufficient increase in accessibility to be affected by the new infrastructure. Only on the right end of the plot, for urban zones close enough to Metro C, we capture a statistically significant increase in economic activities with respect to the baseline. Urban zones on the right of the grey dotted line are the treated ones, on the left are the control ones. The threshold value, chosen as an ansatz in the beginning of the work, takes its justification here, since the regression is significantly different from zero for values of acces-sibility greater than the threshold.

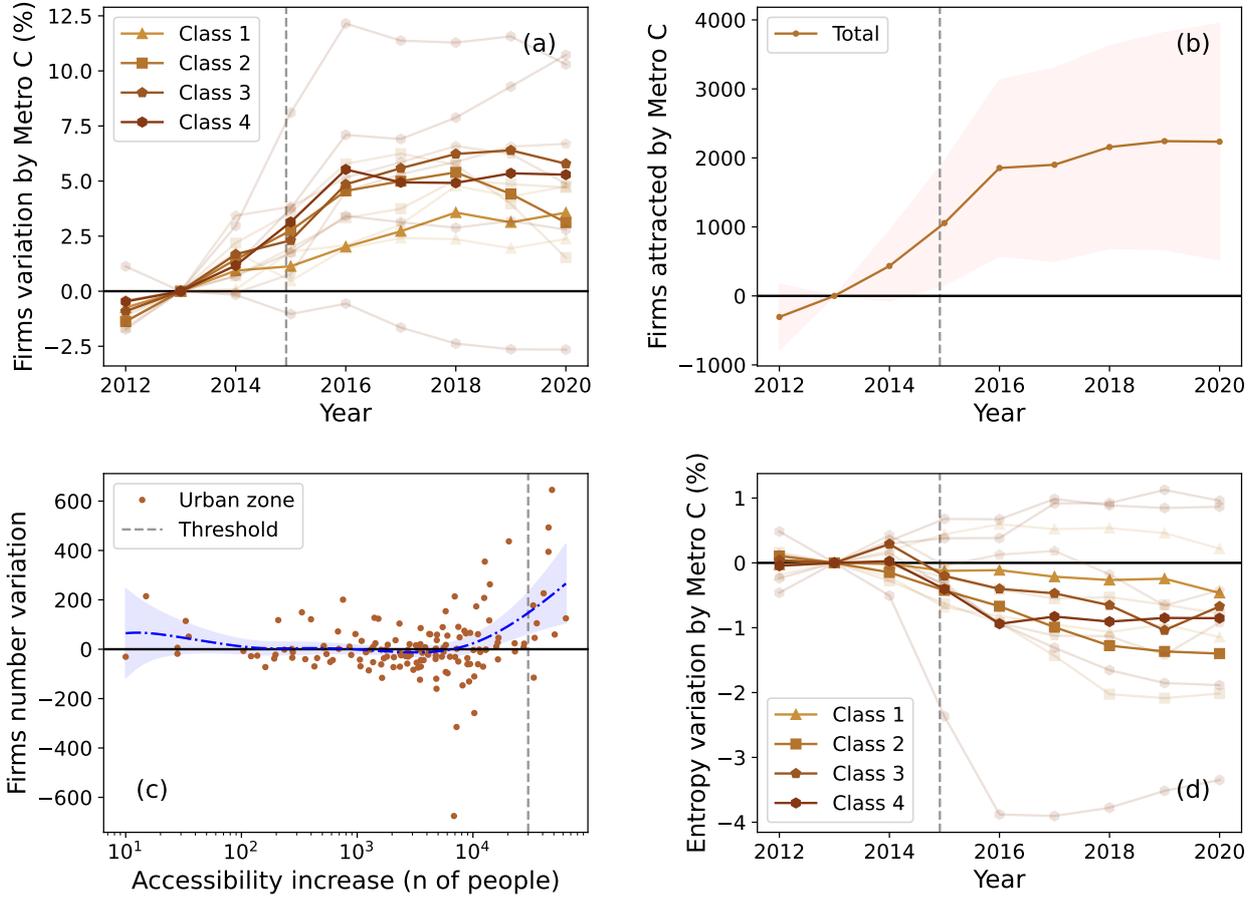

Figure 2: **DID estimation of the effect of accessibility increase on the number of activities in treated areas.** Panel (a) shows the enhancement in the economy of urban zones of Rome interested by the opening of a new metro line with respect to the control group: relative difference in the average number of economic activities per urban zone between zones interested by a new metro opening and zones not touched by it, as a function of the number of years after the openings, for each class of density of activities. Class 4 (the areas with greatest density of economic activities) seems to be the most affected. Transparent curves represent each individual urban zone. In panel (b), global absolute number of economic activities affected by the Metro C opening, computed as the sum of the residuals of each urban zone affected by the new metro with respect to its appropriate baseline. The shadowed region spans a one standard deviation confidence interval. In panel (c), the relationship between the number variation of economic activities due to the introduction of Metro C and the accessibility increase brought by it, for urban zones of Rome. Each dot represents an urban zone, the blue line is a kernel non-parametric regression, the shadowed region corresponds to a 95% confidence interval on the regression, and treated zones occupy the region of the plot on the right of the dashed grey line, while control zones are on the left. The accessibility increase is quantified as the increase in the number of people reachable by public transport from each area due to the new metro line. In panel (d), the mean relative variation in entropy per urban zone is due to the opening of line C of the metro. Transparent curves represent each individual urban zone.

In Fig. 3a, the variation in the number of firms in 2018 relative to the baseline—corresponding to the vertical axis of Fig. 2c—is shown on a map of Rome. Urban zones near Metro C, whose stations are marked with blue dots, tend to appear in red, indicating an increase in the number of economic activities compared to the baseline.

## DID estimation of the impact of Metro C on the diversification of economic activities

The majority of economic activities in the dataset are classified by ATECO codes, an Italian standard representing industry categories. To measure the diversity of the activities inside a certain urban zone, we calculate the Shannon entropy $H(i,t)$, where $i$ represents an urban zone and $t$ a year. The higher the Shannon entropy of a zone in a certain year is, the more diverse are the economic activities of the zone in terms of what it produces or which services it provides. The trend consistently shows a decrease in diversification over time, which tends to become steeper as the economic class ranges from the least developed to the most developed (Fig. 2d). Except for class 1, where the trend of zones with a new metro is totally compatible with the baseline, zones with a new metro tend to decrease their diversity of economic activities with respect to the baseline. This plot suggests that the new metro line encourages specialization in certain economic sectors within urban zones that have a sufficient density of economic activities. Combined with the enhancement in the number of economic activities, this information suggests the formation of clusters of economic activities of the same field after the introduction of the metro line. A similar trend was found in [13] while assessing the impact on firms distribution after the introduction of a new metro line in the suburbs of Madrid, called Metrosur. A possible explanation is that since the new infrastructures allow for faster long trips, the geographical proximity of different kind of activities is less needed than without the metro.

To understand the composition of these clusters, we examined which types of economic activities appear to have benefited from the metro's development and which may have been negatively affected. Table 1 shows the sectors that were favoured most by the introduction of the metro across firm density classes, while Table 2 lists those most disadvantaged. Construction works are favoured by the metro regardless of the class the zone belongs to, frequently in addition to civil engineering offices. This same pattern is also observed in the studies examining the impact of Metrosur in Madrid [13] and the new metro line in Urumqi [15]. They may be interpreted as indicators of transit-oriented development. The retail commerce sector is also favoured by the metro, regardless of the density of pre-existing economic activities.

This same phenomenon is once again seen in Madrid and Urumqi [15, 13] too. Specifically, the former of these case studies finds a difference between these two sectors: while construction works firms are favoured on the whole area of the municipality connected by Metrosur, retail activities tend to cluster in the surroundings of transport stations, decreasing their density on zones away from the new stations. These two mechanisms could be also operating on the zones of Rome connected by Metro C, but our discretization of space is not fine grained enough to capture this diversity. Finally the introduction of the metro in zones already dense in economic activities (i.e. belonging to class 4) pushes a specialization of those areas towards high complexity fields, with these fields being: support to economic activities, personal service activities, advertising activities.

On the other hand, sectors disadvantaged by the introduction of the metro follow a different pattern. There is no sector which is disadvantaged by the metro in zones of every class. This suggests that the decrease in number of activities of a certain type with respect to control zones was not due to a general negative influence of the metro on some sectors, but to the formation of spatially separated clusters of economic activities of the same type.

In Fig. 3b, the relative entropy variation in 2018 is shown with a colour code on a map of Rome. Urban zones close to Metro C, whose stations are marked as blue dots, tend to be blue, meaning they decrease their diversity of economic activities with respect to the baseline.

| Rank | Class 1 | Class 2 | Class 3 | Class 4 |
|------|---------|---------|---------|---------|
| 1 | civil engineering | telecommunications | retail | support to enterprises |
| 2 | construction works | civil engineering | construction works | retail |
| 3 | car and motorcycles commerce | construction works | support to enterprises | construction works |
| 4 | retail | retail | fabrication of metal products | personal service activities |
| 5 | bets | wholesale | telecommunications | advertising |

Table 1: Sectors favoured by the metro, with relative variation of enterprises number in zones affected by metro C over the 7 years between 2013 and 2020

| Rank | Class 1 | Class 2 | Class 3 | Class 4 |
|------|---------|---------|---------|---------|
| 1 | clothing | restaurants | manufacturing | lodging |
| 2 | finance and assurance | scientific and technical activities | clothing | transport |
| 3 | transport | company management | waste collection and disposal | finance and assurance |
| 4 | manufacturing | leather articles | bets | metal products |
| 5 | waste collection and disposal | information | storage | company management |

Table 2: Sectors disadvantaged by the metro, with relative variation of enterprises number in zones affected by metro C over the 7 years between 2013 and 2020

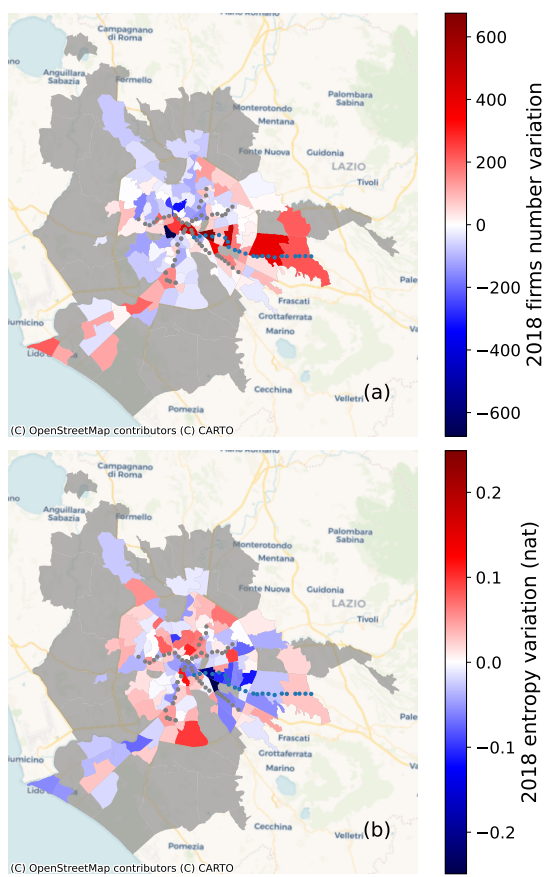

Figure 3: **Spatial variations in the number and diversity of firms.** Panel (a) shows the variation in the *number* of economic activities across urban zones in Rome in 2018, relative to the baseline trend. Panel (b) displays the variation in the *diversity* of economic activities, also relative to the baseline trend. Diversity is quantified using Shannon entropy.

## Impact on micro-enterprises

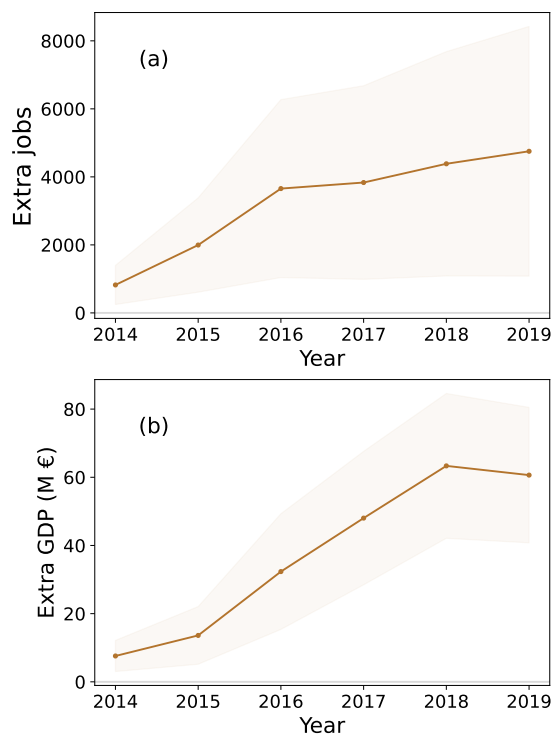

Figure 4: **Effects of the introduction of Metro C on micro-enterprises.** In (a), temporal evolution of the number of jobs in micro-enterprises attracted in treated zones by the introduction of Metro C. In (b), temporal evolution of the variation in GDP in micro-enterprises due to the introduction of Metro C, in treated zones. The shadowed regions correspond to one standard deviation.

An economic activity qualifies as a *micro-enterprise* if its number of employees and profits fall below specific thresholds. Small activities like bars, restaurants, bakeries, laundries, small shops are usually micro-enterprises. Micro-enterprises make up 90% of the dataset under study.

We evaluate the impact of Metro C on the number of jobs and the gross domestic product (GDP) generated by micro-enterprises attracted to the areas served by Metro C. Figure 4a shows the yearly cumulative number of jobs in these micro-enterprises across all treated zones, while Figure 4b displays the total variation in GDP generated by the same micro-enterprises. Both indicators show a marked upward trend over the observed period. See the Supplementary Information for detailed estimates of new jobs and extra GDP in each urban zone.

# Discussion and conclusions

With this study, we find that the opening of Metro C in Rome was linked to beneficial effects on the economic development of the zones it connected, providing better accessibility to local economic activities. We find a statistically relevant increase in the number of economic activities in zones of the city newly connected by Metro C. We also find a mild decrease in the diversity of economic activities in those zones, which we interpret as a hint of clustering of activities of the same kind. In micro-enterprises, we highlight an increase in jobs and GDP produced.

The methodology used to uncover these dynamics is a tailored difference-in-differences approach, which considers a multiplicative effect of the infrastructure on the number and diversity of economic activities, instead of an additive one. This hypothesis is motivated by a lognormal distribution of the number of activities in urban areas, which is the signature of underlying multiplicative processes. A consequence of this approach is to expect equal effects in relative size, so more relevant absolute variations on more developed areas or on bigger ones. To extract the so-called treated group of urban areas, i.e. those affected by the new metro line, we calculated the extra amount of citizens that could reach each urban area of the city with a journey of average length by public transport after the Metro C introduction. We consider as treated zones those areas where this accessibility indicator increased the most.

Treated areas show an increase in economic activities, with respect to other zones with similar economic conditions which provide a reference control group. This trend seems to be starting before the opening of the new metro line, in line with previous findings reported in literature. For instance, a study on whether premiums on housing prices were paid for the expected benefits offered by a new tunnel before its completion [21] found that there were positive price expectation effects well before the completion of the tunnel. Another work analysed the changes in housing prices in areas around a new metro station in Santiago [22]. The average apartment price rose after construction of the new metro line was announced and also at the moment information on the basic engineering project was unveiled, identifying the location of the future stations, before the actual inauguration of the infrastructure. The effect we find is continuing until after the metro line is inaugurated, with firms attracted until 2016, when a more stable phase sets in (Fig. 2).

We also find hints of a decrease in diversity (measured with Shannon entropy) in treated areas relative to the control group. An urban area containing bookshops, restaurants and dentists is more diverse than one containing only cafés. This can be read as a mild effect of clusterization of activities belonging to the same category driven by the increased mobility provided by the new metro. If moving becomes easier, the need to have all services in the same neighbourhood, in a 15-minute fashion, could become less pressing.

Focusing on micro-enterprises, always with a DID estimation, we also highlight an increase in GDP produced and in jobs available in treated urban areas, subsequent to the construction of the metro. This pattern suggests that micro-enterprises in Rome did not encounter the same challenges faced by small businesses during the construction phase of a new light rail line in Charlotte, North Carolina [23]. Charlotte is a more car-dependent urban environment due to its low-density and sprawled suburban development [24], which makes it more susceptible to perturbations in car traffic caused by construction works.

As it can be seen in Fig. (2a), there is one unique zone which is not positively affected by the metro C introduction. This is Tuscolano Nord, the most central area among the affected ones. Even if the number of citizens that can access this urban zone with an average public transport journey grew, the *relative* variation of this quantity was the smallest over treated zones: the area was already well served by public transport, and the introduction of Metro C did not bring a variation in accessibility as important as in the other urban areas.

A potential downside of introducing new infrastructures into established urban areas is the risk of gentrification. Wealthier households may move into neighbourhoods affected by the new development, displacing lower-income residents, and therefore preventing them from benefiting from the new infrastructure.

Generally, empirical evidence in the literature supports the contention that public transport investments tend to increase the value of nearby residential properties [25, 26, 27, 28]. In contrast, Daniele et al. [12] found no such effect in areas affected by the opening of Metro C. No statistically significant variations in house prices were observed in central areas. Conversely, a negative and statistically significant impact on average house prices was detected in peripheral areas, driven by higher-value properties. The authors suggest that poorer households may have perceived the metro as an amenity, while wealthier residents, more reliant on cars, might have viewed it as a disamenity. This conjecture is in line both with the results of another case study on a suburb of Washington, DC [29] and with the meaning that car drivers and transit riders can attach to mobility, reflecting a form of class-based segregation [30]. These findings suggest that this public transport infrastructure was, at least partially, able to address transport poverty.

One limitation of the difference-in-difference approach employed here lies in the possible non-random placement of the infrastructure whose effect is studied. For example, local governments may decide to build a new metro line through a declining neighbourhood to revitalize it, or through a booming area to support its continued growth. A difference-in-difference estimation, which compares the impacted neighbourhood before and after the opening of the new metro line to non-impacted neighbourhoods, can yield biased estimates if the area was already on a diverging economic trajectory prior to the intervention. To address this challenge, we performed a placebo test, using preschools instead of economic activities and applying the same methodology. The impact of Metro C shows to be not statistically relevant in the period analysed, as we expected in case the treated zones were not planned for development or conversely experiencing a decay in population and attractiveness with respect to control areas. In other words, the placebo test corroborated the parallel trends assumption necessary to apply the DID methodology.

With this study we uncovered the implications the construction of Metro C had on the economic fabric of Rome. More in general, we shed light on the general effect that a new public transport infrastructure can have on the areas it connects, through the relationship between the increase in transport access given by the new infrastructure, and their increase in number of economic activities, depicted in Fig. (2c). Through a case study, this work provides insights into the positive spillover effects that can be expected from public transport infrastructure expansions, an important consideration in the broader effort toward achieving carbon neutrality.

# Methods

We discretized the area of Rome into the so-called "Zone Urbanistiche" [31], 155 areas established in 1977 on the basis of a criterion of urbanistic unity, which serve as the basic reference units for urban planning and management activities. In the text, we referred to them as "urban zones".

## Accessibility increase due to the introduction of Metro C

To quantify the accessibility variation due to the introduction of Metro C, we employ a methodology originally defined in [19]. We first cover the city with a hexagonal grid of side 200 m, and compute the travel times by foot between all pairs of hexagon centroids, using the Open Source Routing Machine (OSRM) [32] and OpenStreetMap data [33], assuming a walking speed of 5 Km/h. We then run the Connection Scan Algorithm [34] over the public transport schedule provided by Roma Servizi per la Mobilità [35] and the foot travel times, obtaining the travel times with public transportation under the assumptions that a) the fastest route between two centroids is the preferable one and b) travellers can walk for up to 15 minutes to switch between different public transport lines, to reach the first public transport stop of their journey, and to move from the last stop to their destination. Since the frequency and availability of public transport changes throughout the day, we repeated the calculations for starting times $t_0$ from 6AM to 10PM, at intervals of one hour. We can now define $A_\lambda(\tau, t_0)$ as the area from which centroid $\lambda$ is reachable through a journey with duration $\tau$ starting at time $t_0$, and $P_\lambda(\tau, t_0)$ is the population residing within $A_\lambda(\tau, t_0)$. An indicator of $\lambda$'s accessibility can then be obtained by averaging over all starting and travel times:

$$s_\lambda = \left\langle \int_0^\infty P(\tau, t_0) f(\tau) \, d\tau \right\rangle_{t_0} \quad (1)$$

where $f(\tau)$ is the probability that a trip with public transport has duration $\tau$ [36]. This quantity can be interpreted as the average number of people who can reach $\lambda$ during a typical day. If we define $s'_\lambda$ as the same quantity calculated after removing Metro C from the public transport schedule, we can finally define the quantity plotted in Fig. 1 as

$$\Delta s_{\text{zone}} = \frac{\sum_{\lambda \in \text{zone}} (s_\lambda - s'_\lambda) \, p_\lambda}{\sum_{\lambda \in \text{zone}} p_\lambda} \quad (2)$$

where $p_\lambda$ is the population residing in the hexagon with centroid $\lambda$ and the average is computed over all hexagons whose centroids fall into a given urban zone.

## Data

### Firms' dataset

The dataset about economic activities comes from Infocamere. It contains information on the economic activities open in Rome from 2012 to 2021. It lists 603 624 economic activities in the municipality of Rome, their opening and possibly closing date, their location, and in which economic sector they operate, this last information encoded by an ATECO code. The opening dates range from 1680/12/8 to

2020/8/28 and the closing ones from 2010/1/1 to 2020/12/16.

The dataset presents an outlier: half (51%) of the economic activities located in the urban zone of Torrespaccata abruptly close in 2015 (805 economic activities over 1582). We could not find any reasonable explanation for such behaviour and therefore we treated data of this urban zone as an outlier and excluded them from our calculations.

**Schools' dataset**

Data on preschools are provided by the municipality of Rome and are publicly available [37].

## Difference-in-differences impact estimations

Usually difference-in-differences estimators are derived using a linear parametric model [18]. The outcome variable $Y$ of unit $i$ at time $t$ is written as [38]

$$Y_i(t) = \hat{\delta}(t) + \hat{\alpha}(t) D(i,t) + \hat{\eta}_i(t) + \hat{v}_i(t), \quad (3)$$

where where $\hat{\delta}(t)$ is a function of time only, modelling a trend independent on the unit $i$ considered, $\hat{\alpha}(t)$ represents the impact of the treatment, $\hat{\eta}_i(t)$ is an individual-specific component, $\hat{v}_i(t)$ is an individual-transitory shock that has mean zero at each period, $t < t^*$ and $t > t^*$, and is possibly correlated in time [18], $t^*$ is the time of onset of the treatment, the year of opening of the new metro in our case, and finally $D(i,t)$ is a dummy taking value 1 for treated zones after the treatment onset and zero otherwise; therefore it is defined as

$$D(i,t) = \mathcal{H}(t - t^*) \prod_{j \in T} \delta_{ij}, \quad (4)$$

where $\mathcal{H}$ is the Heaviside function, $\delta_{ij}$ the Kronecker delta, and $T$ is the set of treated units.

If one then considers two instants,

$$t_{\rm b} < t^* \text{ and } t_{\rm a} > t^*$$

(where b and a stand for before and after), the impact of the treatment can be written as

$$\hat{\alpha} = \{E[Y_i(t_{\rm a})|D(i,t_{\rm a})=1] - E[Y(i,t_{\rm a})|D(i,t_{\rm a})=0]\} - \\ \{E[Y_i(t_{\rm b})|D(i,t_{\rm b})=1] - E[Y(i,t_{\rm b})|D(i,t_{\rm b})=0]\},$$

which makes $\hat{\alpha}$ estimable by least squares regression of $Y$ before and after treatment. For details see [18].

Here instead we consider a multiplicative effect of the intervention on the outcome variable $Y$. The motivation for this modelling lays in the law of proportionate effect [39]. A process, described by a variable $X$ at discrete time intervals $t_k$, is said to obey the law of proportionate effect if [40]

$$X_{t_k} = X_{t_{k-1}} (1 + \epsilon_k), \quad (5)$$

where each $\epsilon_k$ is statistically independent of $\{X_{t_k}\}$ and is extracted independently from a unique distribution. To define $X_{t_n}$ we then need a set $\{\epsilon_k\}_{k=1\ldots n}$ of mutually independent and identically distributed random variables. If the distribution of the $\epsilon_k$ has finite mean and variance, different realizations of $X_{t_n}$, with different sets of $\{\epsilon_k\}_{k=1\ldots n}$ drawn from the same distribution, are (asymptotically for large $n$) lognormally distributed [40].

We observe in our data that the distribution of the number of economic activities per urban zone is lognormal. As an example, in the last year of our dataset, 2021, the Kolmogorov-Smirnov test on the logarithms of the numbers of economic activities per urban zone returned a statistic of 0.094, making their distribution undistinguishable from a Gaussian at 95% C.L.. The distribution is shown in SI. We then argue that the process giving rise to this distribution if of the form of Eq. (5), and therefore time evolutions of the quantities obey a multiplicative process. DID estimations can be performed also for this kind of processes [41]. In this view, the quantity $X$ is our outcome variable $Y$ and different urban zones are described by a different set of $\{\epsilon_k\}_{k=1\ldots n}$, depending on their specific characteristics. If the effect of such characteristics on the outcome variable depends multiplicatively on the previous value of the variable, a better writing of our outcome variable than Eq. (3) is

$$Y_i(t) = [1 + \alpha(t) D(i,t)] \delta(t) \eta_i(t). \quad (6)$$

Therefore, after $t^*$ in the treated zone $i$, there's an extra contribution to the outcome variable, due to the intervention, which is, dropping the time dependence for ease of notation, $\Delta Y_i = \alpha \delta \eta_i$. Let us now average the outcome variable separately over treated zones

$$\bar{Y}_T(t) = \frac{1}{|T|} \sum_{i \in T} Y_i(t)$$

and over untreated ones

$$\bar{Y}_U(t) = \frac{1}{|T^C|} \sum_{i \notin T} Y_i(t),$$

with $|T|$ representing the cardinality of $T$, and $T^C$ the complementary of $T$. Then, leaving again implicit the dependence on time of quantities,

$$\bar{Y}_T = (1 + \alpha) \delta \bar{\eta}_T,$$

where $\bar{\eta}_T$ is the average of $\eta_i$ over treated units, and analogously

$$\bar{Y}_U(t) = (1 + \alpha) \delta \bar{\eta}_U.$$

Via simple algebraical operations we then find that the average extra contribution to $Y$ due to the intervention, on treated zones, is

$$\Delta Y_T = \bar{Y}_T(t) - \frac{\bar{Y}_T(t^*)}{\bar{Y}_U(t^*)} \bar{Y}_U(t) = \\ = \bar{Y}_T(t) - Y_{\rm baseline}(t), \quad (7)$$

where in the last row we just defined $Y_{\text{baseline}}(t)$ as the second term of the first row. In fact, the first term defines the orange curves of Supplementary Figure 6 and 7, while the second term defines the gray baseline.

For the methodology being meaningful the general trend $\delta$ and the multiplicative effect of the intervention $\alpha$ must not vary over treated and untreated zones. To ensure this, we divided our units, the urban zones of Rome, in quintiles based on their density of economic activities. The quintile with the least density does not host any treated zone, the remaining four are used in the study. All the methodology explained has thus to be intended inside each quintile separately.

### Identification of treated and control zones

We classify a urban zone as "treated" if after the construction of Metro C it could be easily accessed by an additional number of people exceeding a critical value. This critical value is set to 30 000 people, which is approximately half of the maximum increase in accessibility observed for any urban zone, the latter being 63 000 people. Zones that do not fall into the treated group provide the "control" group.

### Testing the parallel trends assumption

A key identifying assumption of the difference-in-differences approach is that, in the absence of treatment, treated and control units would have followed parallel trends over time. To provide empirical support for this assumption in the context of our study, we conducted a placebo analysis using a sample of preschools (early childhood education centres for children aged 3–5) instead of firms.

On the time scale of our analysis, preschools are unlikely to be directly affected by the new Metro C line, nor are they expected to systematically respond to the kind of urban development policies that might accompany such infrastructure investments. If the treated areas were already on a distinct development trajectory prior to the metro project—for instance, due to a broader urban plan—it would be reasonable to expect significant effects even in this placebo sample.

However, our DID estimation on preschools yields no statistically significant results. Specifically, the estimated effect of Metro C on the number of preschools in treated urban zones in 2018 is $0 \pm 4$, expressed as the average change per urban zone. This absence of effect suggests that treated neighborhoods were not subject to independent development efforts prior to the intervention. Overall, these findings support the validity of the parallel trends assumption for our main analysis.

An event-study figure analogous to Fig. 2, based on the preschool data, is provided in SI. Additionally, the year-by-year variation in the number of preschools attributed to Metro C by urban zone is reported in tabular form.

### DID estimation of the impact of Metro C on the number of economic activities

In order to have a synthetic picture of the effect of the opening of the metro line on the economy of the city, the relative increment $\Delta_R n_T$ in the number of enterprises of each year $t$ and each zone $i$ affected by the metro relative to its appropriate baseline have been computed as follows:

$$\Delta_R n_T = \frac{\Delta n_T}{n_{\text{baseline}}} \times 100\% \\ = \frac{\bar{n}_T - n_{\text{baseline}}}{n_{\text{baseline}}} \times 100\%, \quad (8)$$

where $\Delta n_T$ is the quantity $\Delta Y_T$ defined in the previous paragraph in the case in which the generic outcome variable $Y$ is the number of economic activities $n$ per the urban zone. $\bar{n}_T$ is the average number of economic activities in treated zones and $n_{\text{baseline}}$ is the baseline, i.e. the normalized average over zones not interested by the metro opening. All the quantities are functions of time and they refer to averages and baselines of zones belonging to the same economic class, defined on the basis of the density of economic activities. Then $\Delta_R n_T$ defines the mean effect of the introduction of Metro C on the number of economic activities per urban zone. It is shown, for each economic class, over six years, in Fig. 2a.

The global number of economic activities attracted by Metro C, shown in Fig. 2b, has been computed simply summing all economic activities attracted by the metro in all treated zones, as follows:

$$\Delta n_{T,\,\text{tot}} = \sum_{i \in T} \Delta n_{T,\,i}. \quad (9)$$

The kernel non-parametric regression of the economic activity number variation with respect to the baseline as a function of the accessibility increase brought by Metro C, depicted in Fig. 2c, has been computed with a local linear estimator, a gaussian kernel and bandwidth selected using the AIC Hurvich bandwidth estimation. The 95% C.I. was estimated via bootstrapping with 2000 iterations.

### DID estimation of the impact of Metro C on the diversification of economic activities

The majority (87%) of the economic activities listed in the dataset under consideration is marked by at least one ATECO code, defining what the economic activity produces or in which sector it operates [42]. These codes have a hierarchical structure: the first two digits of them define the most coarse grained classification of economic activities, while each one of the following four digit refers to subcategories.

For each urban zone $i$ and for each year we computed the Shannon entropy $H$ of economic activities, defined as

$$H(i,t) = \sum_j f_j(i,t) \, \ln f_j(i,t) \,, \qquad (10)$$

where the index $j$ runs over the ATECO codes and $f_j(i,t)$ is the ratio between the number of economic activities operating in the sector corresponding to the ATECO code $j$ in a certain zone $i$ at year $t$, over the total number of economic activities open in the zone $i$ in the year $t$:

$$f_j(i,t) = \frac{n_j(i,t)}{\sum_k n_k(i,t)} \,. \qquad (11)$$

The relative variation shown in Fig. 2d has been computed as an adjusted version of the relative firms number variation (8):

$$\Delta_R H_T := \frac{H_T - H_{\text{baseline}}}{H_{\text{baseline}}} \times 100\% \,. \qquad (12)$$

To study which kind of economic activities are favoured and which are disadvantaged by the opening of the metro, we assessed the impact of the new metro on the distribution of activities in each sector, and for each class separately. We calculated the difference $d$ between from one side the number of economic activities open for each ATECO code 6 years after the opening of the metro, and to the other side the same quantity calculated one year before the opening of the metro. In Supplementary Figures from 8 to 11 the mean of such a difference for treated zones, in orange, is shown with the same quantity for control zones, in grey.

If we call $d_T$ this difference of economic activities before and after the introduction of the metro in treated zones, and $d_{\text{baseline}}$ the one of the baseline, we can define the quantity

$$\Delta := \frac{d_{\text{metro}} - d_{\text{baseline}}}{\sqrt{|d_{\text{baseline}}|}} \,,$$

which represents a difference between the two distributions, for treated and control zones, normalized to the statistical error of the baseline, supposing a Poissonian distribution for the number of enterprises of each ATECO code.

In SI, this measure of the difference between new metro areas and the control group, and therefore of the effect of the introduction of the metro, is plotted as a function of the ATECO code for each economic class.

We can then rank the ATECO codes on the basis of their value of the relative increment $\Delta$ due to the metro. The first 5 sectors of this rank, i.e. the most favoured by the introduction of the metro, are listed for every economic class, in Table 1; the last 5 of the ranking, i.e. the sectors most disadvantaged by the introduction of the metro, are on the other hand listed on Table 2.

## Impact on micro-enterprises

A micro-enterprise can be defined as a firm having less than 10 employees and an annual turnover (the amount of money taken in a particular period) or balance sheet (a statement of a company's assets and liabilities) below 2 million of euros [43]. We assess the impact the line C of the metro on micro-enterprises, in particular assessing the number of jobs and the extra GDP created as a consequence of the introduction of such infrastructure. We quantify the variation $\delta n$ in micro-enterprises due to the introduction of the metro, per year per urban zone, following the same difference-in-differences technique. The results for each urban zone affected by the metro are collected in the Supplementary Information.

**Variation in the number of jobs induced by the introduction of Metro C** We call $T$ the set of treated urban zones, and we call $\mathcal{A}_i$ the set of ATECO codes present in zone $i$, meaning that for each code in $\mathcal{A}_i$ it exists in zone $i$ at least one economic activity labelled with that code. With this notation the number of jobs $\delta W$ created or displaced as a consequence of the introduction of Metro C can be calculated as follows:

$$\delta W(t) = \sum_{i \in T} \sum_{j \in \mathcal{A}_i} f_j(i,t)\, \delta n(i,t)\, e_j(i,t) \,, \qquad (13)$$

where $f_{ij}(t)$ is the frequency of the ATECO code $j$ in the micro economic activities of zone $i$ opened from 2012 till the year $t$, normalized in such a way that

$$\sum_{j \in \mathcal{A}_i} f_j(i) = 1 \,,$$

$\delta n(i,t)$ is the variation due to the metro of the number of micro economic activities in the urban zone $i$ in the year $t$, and $e_j(i,t)$ is the average number of employees of micro economic activities of type $j$, in the zone $i$ in the year $t$. Therefore the quantity $f_j(i,t)\,\delta n(i,t)$ estimates the variation in economic activities of type $j$ in the zone $i$ at year $t$; multiplying if for $d_j(i,t)$ one obtains the variation in jobs due to the metro in micro economic activities of ateco code $j$ in the zone $i$, and summing over all $j$s and $i$s we obtain the total impact of the metro in terms of jobs. The variations in the number of jobs caused by the introduction of the metro in a zone $i$ in a year $t$, for $i$ in $T$ and $t$ between 2014 and 2019 are collected in Supplementary Information.

**Variation in GDP induced by the introduction of Metro C** To evaluate the variation in GDP caused by Metro C, it can be used the same approach applied for the variation in jobs: it is sufficient to replace the average number of employees in micro economic activities of type $j$ in the zone $i$ with the corresponding mean added value $v_j(i)$. Therefore the variation in GDP induced by the introduction of the metro on zone $i$ by micro economic activities of type $j$ is given by $\delta\text{GDP}_j := f_j(i,t)\,\delta n(i,t)\,v_j(i,t)$, and by summing over all the ATECO codes we get the extra GDP due to the metro for each zone and each year.

If we then sum over all the treated zones we can estimate the total variation in gross domestic product $\delta\text{GDP}$ due to Metro C. Therefore we obtain:

$$\delta\text{GDP}(t) = \sum_{i \in T} \sum_{j \in \mathcal{A}_i} f_j(i,t)\,\delta n(i,t)\,v_j(i,t)\,. \quad (14)$$

## Acknowledgments

The authors would like to thank Elsa Arcaute, Rossano Schifanella and Bernardo Monechi for their enlightening comments. Hygor P. M. Melo acknowledges the support of Fundação Edson Queiroz, Universidade de Fortaleza, and Fundação Cearense de Apoio ao Desenvolvimento Científico e Tecnológico.

## Data availability

The population distribution data used can be downloaded from WorldPop [44]. The boundaries of urban areas can be sourced from OECD [45], and, in instances where OECD data are unavailable, from the Global Human Settlement files [46]. The EDGAR dataset is also publicly available [47]. To compute proximity times, POIs data are publicly available on OpenStreetMap [33], while travel times were calculated using the Open Source Routing Machine (OSRM) [48]. Firms' data used in this study have been provided by Infocamere and cannot be publicly shared in raw format. The rest of the data employed here is publicly available and can be provided upon request.

## Code availability

The code used for the analysis and visualizations is available from the corresponding author upon reasonable request.

## Author Contributions

Research design and study concept: F.M., H.P.M.M., M.B., V.L.. Data analysis: F.M., B.C.. Result interpretation: F.M., B.C., H.P.M.M., M.B., V.L.. Manuscript drafting: F.M.. Manuscript review and editing: F.M., B.C., H.P.M.M., M.B., V.L.. All authors have read and approved the manuscript.

## Competing Interests

The authors declare no competing interests.

## Additional Information

Ethics, Consent to Participate, and Consent to Publish declarations: not applicable.
Funding: not applicable.
Clinical trial number: not applicable.

## References

[1] *World Energy Outlook*. Tech. rep. IEA, 2024. URL: http://www.iea.org.

[2] S. Bouckaert A. F. Pales C. McGlade U. Remme B. Wanner L. Varro D. D'Ambrosio T. Spencer. *Net Zero by 2050: A Roadmap for the Global Energy Sector*. Tech. rep. 9, Rue de la Federation, Paris 75015, France: International Energy Agency, 2021.

[3] UITP. *World Metro Figures 2021: Statistics Brief*. Tech. rep. International Association of Public Transport, May 2022. URL: https://www.uitp.org.

[4] Daniel G. Chatman and Robert B. Noland. "Do Public Transport Improvements Increase Agglomeration Economies? A Review of Literature and an Agenda for Research". In: *Transport Reviews* 31.6 (Nov. 2011), pp. 725–742. ISSN: 0144-1647, 1464-5327. DOI: 10.1080/01441647.2011.587908. (Visited on 08/01/2025).

[5] Alex Anas, Richard Arnott, and Kenneth A Small. "Urban spatial structure". In: *Journal of economic literature* 36.3 (1998), pp. 1426–1464.

[6] Patricia Rice, Anthony J Venables, and Eleonora Patacchini. "Spatial determinants of productivity: Analysis for the regions of Great Britain". In: *Regional science and urban economics* 36.6 (2006), pp. 727–752.

[7] Daniel J Graham. "Agglomeration, Productivity and Transport Investment". In: *Journal of Transport Economics and Policy* 41 ().

[8] Qisheng Pan. "The impacts of an urban light rail system on residential property values: A case study of the Houston METRORail transit line". In: *Transportation Planning and Technology* 36.2 (2013), pp. 145–169. DOI: 10.1080/03081060.2012.743831.

[9] Nick Tsivanidis. "The Aggregate and Distributional Effects of Urban Transit Infrastructure: Evidence from Bogotá's TransMilenio". In: *Econometrica* 87.5 (2019), pp. 1421–1454. DOI: 10.3982/ECTA15343.

[10] Edoardo Zanchini et al. *Rapporto pendolaria 2021*. Legambiente, 2021.

[11] Antonio Accetturo et al. "Sviluppo locale, economie urbane e crescita aggregata (Local Development, Urban Economies and Aggregate Growth)". In: *Bank of Italy Occasional Paper* 490 (2019).

[12] Federica Daniele and Elena Romito. *The impact of Metro C in Rome on the housing market*. Banca d'Italia, 2022.

[13] Lucia Mejia-Dorantes, Antonio Paez, and Jose Manuel Vassallo. "Transportation infrastructure impacts on firm location: the effect of a new metro line in the suburbs of Madrid". In: *Journal of Transport Geography* 22 (2012). Special Section on Rail Transit Systems and High Speed Rail, pp. 236–250. ISSN: 0966-6923. DOI: https://doi.org/10.1016/j.jtrangeo.2011.09.006. URL: https://www.sciencedirect.com/science/article/pii/S0966692311001487.

[14] Csaba G Pogonyi, Daniel J Graham, and Jose M Carbo. "Metros, agglomeration and displacement. Evidence from London". In: *Regional Science and Urban Economics* 90 (2021), p. 103681.

[15] Aishanjiang Abudurexiti, Zulihuma Abulikemu, and Maimaitizunong Keyimu. "POI-Based Assessment of Sustainable Commercial Development: Spatial Distribution Characteristics and Influencing Factors of Commercial Facilities Around Urumqi Metro Line 1 Stations". In: *Sustainability* 17.12 (June 2025), p. 5270. ISSN: 2071-1050. DOI: 10.3390/su17125270. (Visited on 07/30/2025).

[16] Giulio Grossi et al. "Direct and Spillover Effects of a New Tramway Line on the Commercial Vitality of Peripheral Streets: A Synthetic-Control Approach". In: *Journal of the Royal Statistical Society Series A: Statistics in Society* 188.1 (Jan. 2025), pp. 223–240. ISSN: 0964-1998, 1467-985X. DOI: 10.1093/jrsssa/qnae032. (Visited on 07/31/2025).

[17] A. Roukouni, S. Basbas, and A. Kokkalis. "Impacts of a Metro Station to the Land Use and Transport System: The Thessaloniki Metro Case". In: *Procedia - Social and Behavioral Sciences* 48 (2012). Transport Research Arena 2012, pp. 1155–1163. ISSN: 1877-0428. DOI: https://doi.org/10.1016/j.sbspro.2012.06.1091. URL: https://www.sciencedirect.com/science/article/pii/S1877042812028273.

[18] Alberto Abadie. "Semiparametric difference-in-differences estimators". In: *The review of economic studies* 72.1 (2005), pp. 1–19.

[19] Indaco Biazzo, Bernardo Monechi, and Vittorio Loreto. "General scores for accessibility and inequality measures in urban areas". In: *Royal Society open science* 6.8 (2019), p. 190979.

[20] Breed D Meyer. "Natural and quasi-experiments in economics". In: *Journal of business & economic statistics* 13.2 (1995), pp. 151–161.

[21] Chung Yim Yiu and Siu Kei Wong. "The effects of expected transport improvements on housing prices". In: *Urban studies* 42.1 (2005), pp. 113–125.

[22] Claudio A Agostini and Gastón A Palmucci. "The anticipated capitalisation effect of a new metro line on housing prices". In: *Fiscal studies* 29.2 (2008), pp. 233–256.

[23] Sara Tornabene and Isabelle Nilsson. "Rail Transit Investments and Economic Development: Challenges for Small Businesses". In: *Journal of Transport Geography* 94 (June 2021), p. 103087. ISSN: 09666923. DOI: 10.1016/j.jtrangeo.2021.103087. (Visited on 07/31/2025).

[24] Yizhao Yang. "A Tale of Two Cities: Physical Form and Neighborhood Satisfaction in Metropolitan Portland and Charlotte". In: *Journal of the American Planning Association* 74.3 (July 2008), pp. 307–323. ISSN: 0194-4363, 1939-0130. DOI: 10.1080/01944360802215546. (Visited on 07/31/2025).

[25] Edmund Zolnik. "A Longitudinal Analysis of the Effect of Public Rail Infrastructure on Proximate Residential Property Transactions". In: *Urban Studies* 57.8 (June 2020), pp. 1620–1641. ISSN: 0042-0980. DOI: 10.1177/0042098019836564. (Visited on 07/30/2025).

[26] Dongsheng He et al. "New Metro and Housing Price and Rent Premiums: A Natural Experiment in China". In: *Urban Studies* 61.7 (May 2024), pp. 1371–1392. ISSN: 0042-0980. DOI: 10.1177/00420980231208560. (Visited on 07/30/2025).

[27] Miguel Padeiro, Ana Louro, and Nuno Marques da Costa. "Transit-Oriented Development and Gentrification: A Systematic Review". In: *Transport Reviews* 39.6 (Nov. 2019), pp. 733–754. ISSN: 0144-1647. DOI: 10.1080/01441647.2019.1649316. (Visited on 07/31/2025).

[28] Stephen Gibbons and Stephen Machin. "Valuing rail access using transport innovations". In: *Journal of Urban Economics* 57.1 (2005), pp. 148–169. DOI: 10.1016/j.jue.2004.10.002.

[29] Rhea Acuña. "End of the Line: The Impact of New Suburban Rail Stations on Housing Prices". In: *Journal of Transport and Land Use* 16.1 (Feb. 2023), pp. 67–86. ISSN: 1938-7849. DOI: 10.5198/jtlu.2023.2199. (Visited on 07/30/2025).

[30] Matthew Williams and Non Arkaraprasertkul. "Mobility in a Global City: Making Sense of Shanghai's Growing Automobile-Dominated Transport Culture". In: *Urban Studies* 54.10 (Aug. 2017), pp. 2232–2248. ISSN: 0042-0980. DOI: 10.1177/0042098016637568. (Visited on 07/30/2025).

[31] *Delibera consiliare n. 2983*. 29-30 July 1977.

[32] *Project OSRM*. URL: https://project-osrm.org.

[33] *OpenStreetMap*. URL: https://www.openstreetmap.org.

[34] Ben Strasser Julian Dibbelt Thomas Pajor and Dorothea Wagner. "Connection Scan Algorithm". In: *ACM Journal of Experimental Algorithmics* 23.1.7 (Oct. 2018), pp. 1–56. DOI: 10.1145/3274661.

[35] *Roma Servizi per la Mobilità*. URL: https://romamobilita.it/it.

[36] Robert Kölbl and Dirk Helbing. "Energy laws in human travel behaviour". In: *New Journal of Physics* 5.1 (2003), p. 48.

[37] URL: https://dati.comune.roma.it.

[38] Orley C Ashenfelter and David Card. *Using the longitudinal structure of earnings to estimate the effect of training programs*. 1984.

[39] Robert Gibrat. *Les inégalités économiques*. Librairie du Recueil Sirey, Paris, 1931.

[40] Edwin L Crow and Kunio Shimizu. *Lognormal distributions*. Marcel Dekker New York, 1987.

[41] Emanuele Ciani and Paul Fisher. "Dif-in-dif estimators of multiplicative treatment effects". In: *Journal of Econometric Methods* 8.1 (2019), p. 20160011.

[42] *La struttura (codici e titoli) della classificazione delle attività economiche ATECO 2025*. Tech. rep. Nota metodologica, pubblicata l'11 dicembre 2024. ISTAT – Istituto Nazionale di Statistica, Dec. 2024. URL: https://www.istat.it/wp-content/uploads/2024/12/Nota-metodologica-1.pdf.

[43] *Commission Recommendation of 6 May 2003 concerning the definition of micro, small and medium-sized enterprises (Text with EEA relevance) (notified under document number C(2003) 1422)*. May 2003. URL: http://data.europa.eu/eli/reco/2003/361/oj.

[44] *WorldPop*. URL: https://worldpop.org.

[45] URL: https://www.oecd.org/en/data/datasets/oecd-definition-of-cities-and-functional-urban-areas.html.

[46] URL: https://data.jrc.ec.europa.eu/dataset/53473144-b88c-44bc-b4a3-4583ed1f547e.

[47] JRC European Commission. *EDGAR (Emissions Database for Global Atmospheric Research) Community GHG Database (a collaboration between the European Commission, Joint Research Centre (JRC), the International Energy Agency (IEA), and comprising IEA-EDGAR CO2, EDGAR CH4, EDGAR N2O, EDGAR F-GASES version 7.0*. 2022. URL: https://edgar.jrc.ec.europa.eu/dataset_ghg70.

[48] Dennis Luxen and Christian Vetter. "Real-time routing with OpenStreetMap data". In: *Proceedings of the 19th ACM SIGSPATIAL international conference on advances in geographic information systems*. 2011, pp. 513–516.

# Supplementary Information for
# Increasing accessibility by public transport benefits local economy: the effect of a new metro line in Rome

**Accessibility increase due to the introduction of Metro C**

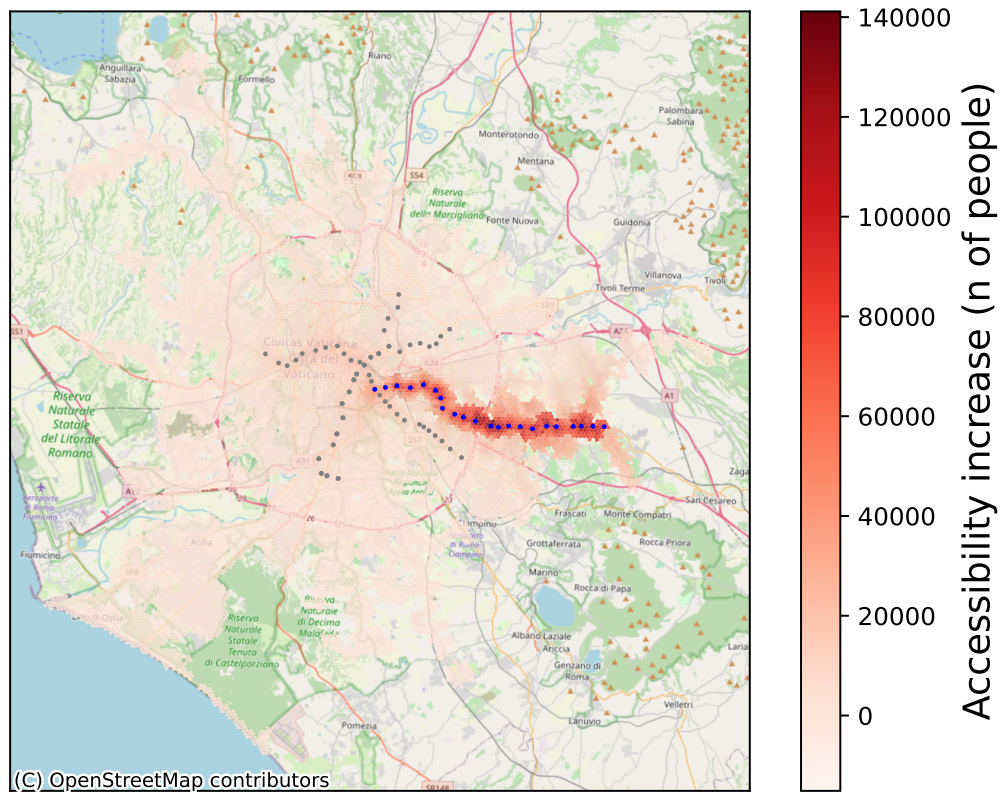

Supplementary Figure 1: Accessibility increase due to the introduction of Metro C, measured as the extra amount of people that can easily reach a given hexagon of the grid shown, with public transport

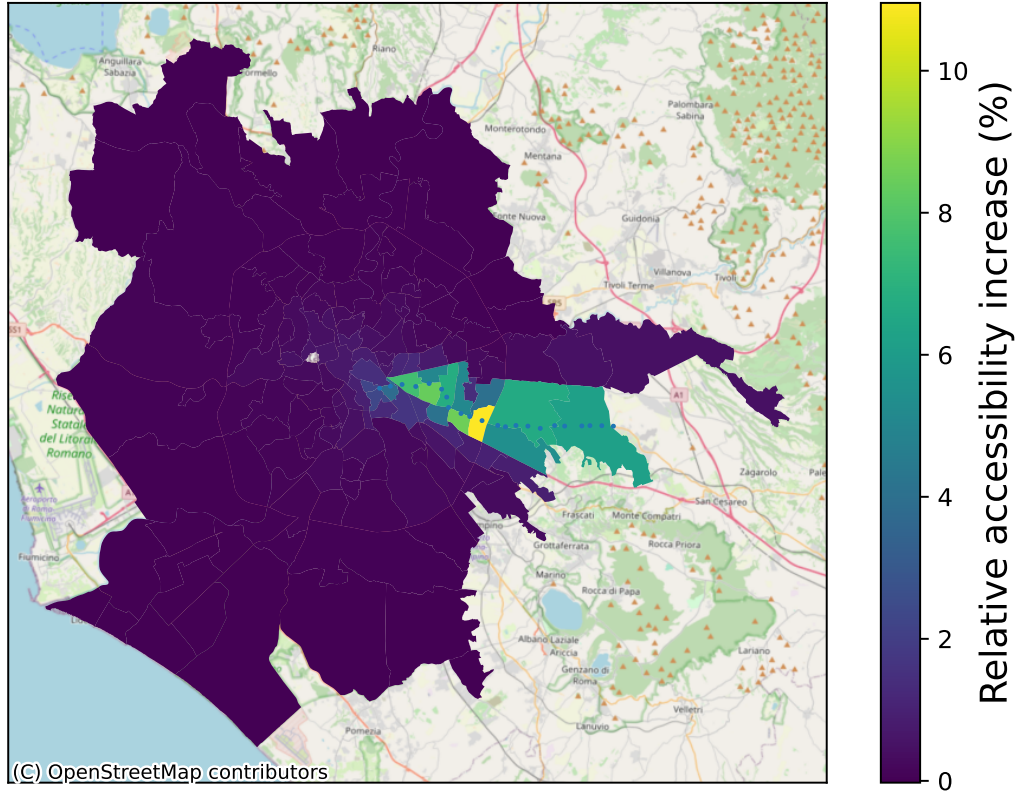

Supplementary Figure 2: Relative accessibility increase due to the introduction of Metro C. Accessibility is measured as the extra amount of people that can easily reach a given hexagon of the grid shown, with public transport. Here it is shown in percent the relative change $(s_\text{f} - s_\text{i})/s_\text{i}$, where $s_\text{i}$ is the initial accessibility, before Metro C, and $s_\text{f}$ is the final one, after the introduction of the metro.

# Diversity/abundance relation of economic activities in urban zones of Rome

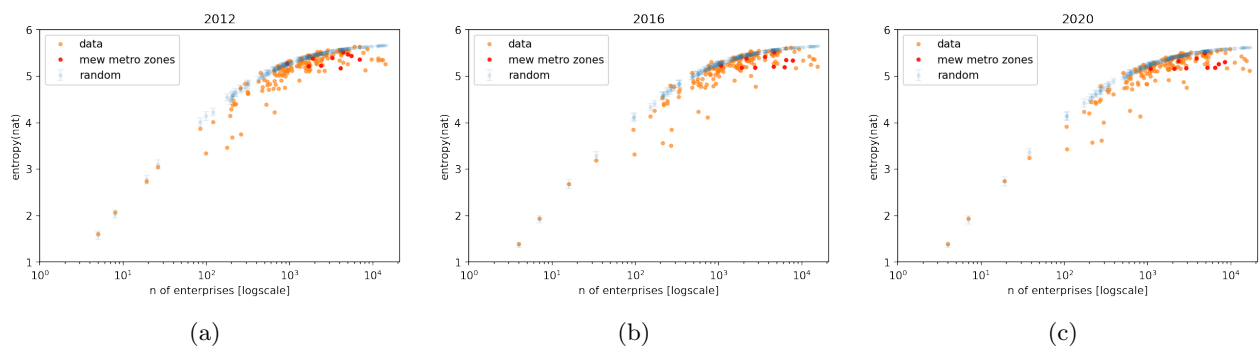

(a)　　　　　　　　　　　　(b)　　　　　　　　　　　　(c)

Supplementary Figure 3: Relation between the abundance of enterprises and their diversification in the urbanistic zones of Rome for different years: 2012, 2016, and 2020, respectively for panels (a), (b) and (c).

The three plots in table 3 inspect the relation between diversity and abundance of the economic activities of Rome: for each plot, each orange or red dot represents an urban zone of the city, in red are zones affected by the introduction of the metro; its abscissa is the number of economic activities present in its territory, and the ordinate is its Shannon entropy $H$, quantifying the degree of diversification of those economic activities. For each abscissa of these datapoints it has also been computed the Shannon entropy of a randomized version of our city, created by shuffling the ateco codes of economic activities in our dataset, while keeping constant the number of economic activities in each urban zone. The mean entropies over a growing number of realizations of the shuffling approach the maximum entropy configuration, which can be computed analytically. The random distribution scenario is represented in the plots by blue points. The three plots refers to three different years, respectively 2012, 2016 and 2020. Over the years one can notice a general tendency of the number of economic activities to grow and polarize and of the zones to specialize. Zones in red, therefore crossed by the new metro, tend to descend in the plot over years faster than the average, and therefore to specialize. There is also a stable pattern throughout the years: the entropy, i.e. the degree of diversification, regardless of the year, initially grows while the abundance of economic activities grows, but then peaks and starts a declining phase, drifting down from the random case which would predict a monotonous increase of entropy with increasing number of economic activities in the zone; in other words for zones very abundant in economic activities, the diversity of economic activities anticorrelates with their number. A similar behaviour is find in ecology, where the diversity of species of a taxon initially grows with increasing *productivity* (which could be taken to be available solar energy, yearly amount of rain, inverse of depth in oceans and so on, depending on the taxon considered), but then peaks and for very high productivity values it declines [1]. In a city, this behaviour can be seen as the result of the formation of clusters of economic activities working in the same field. A more stringent analogy with this behaviour of species can be drawn using as a measure of productivity for economic activities the number of people which can reach the place where the economic activity is located. While this can be a very good proxy for the productivity of small grocery shops and bars, it can be less adequate to estimate the same quantity for very specialized shops or in general activities offering services not widespread in the city. As a first order approximation every economic activity relies on people both as employees and as clients, therefore an accessibility metric can be seen as a proxy for ecological productivity of economic activities.

## Lognormal distribution of the number of firms per urban zone

Figure 4 shows the histogram of the logarithms of the number of economic activities per urban zone of Rome in 2021. Since it is compatible with a normal distribution, the number of economic activities per urban zone is lognormally distributed. This fact motivates our assumption that the increase or decrease in the number of economic activities has to be treated as a multiplicative process. We indeed use this approach in the DiD estimation.

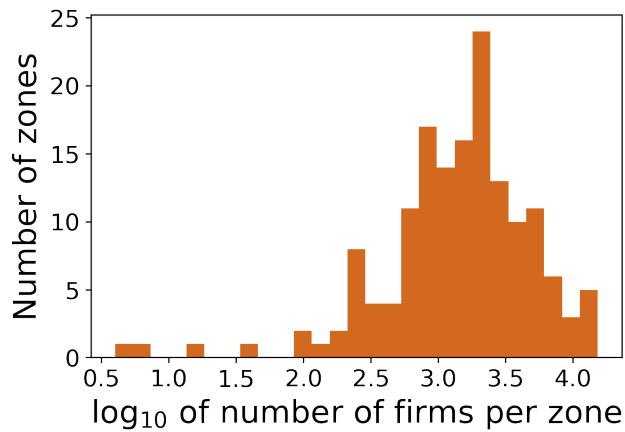

Supplementary Figure 4: Histogram of the logarithms of the number of economic activities per urban zone of Rome in 2021. The distribution is undistinguishable from a Gaussian at 95% C.L.

## Placebo test for the parallel trends assumption

Figure 5 and Table 1 present the results of the DID estimation of the effect of Metro C on preschools, which is not statistically significant. This finding supports the notion that the treated areas were not already slated for development, thereby corroborating the parallel trends assumption, which is fundamental for the validity of the DID approach.

|  | 2011 | 2012 | 2013 | 2017 | 2018 | 2022 |
|---|---|---|---|---|---|---|
| Giardinetti-Tor Vergata | 0.07 | 0.07 | 0.00 | −0.41 | −0.41 | −0.41 |
| Borghesiana | 0.16 | 0.16 | −0.00 | 0.05 | 0.05 | 0.05 |
| Casilino | −0.00 | −0.00 | −0.00 | 0.04 | 0.09 | 0.13 |
| Torre Angela | −0.00 | −0.00 | −0.00 | 0.04 | 0.09 | 0.13 |
| Alessandrina | −0.98 | 0.02 | 0.00 | 0.02 | 0.02 | 0.07 |
| Torre Maura | 0.01 | 0.01 | 0.00 | 0.01 | 0.01 | 0.04 |
| Centocelle | 0.00 | 0.00 | 0.00 | 0.07 | 0.07 | 0.07 |
| Gordiani | −0.00 | −0.00 | −0.00 | 0.04 | 0.04 | 0.04 |
| Tuscolano Nord | 0.00 | 0.00 | 0.00 | 0.06 | 0.06 | 0.06 |
| Torpignattara | 0.00 | 0.00 | 0.00 | 0.06 | 0.06 | 0.06 |

Supplementary Table 1: Estimated variation in the number of preschools due to the introduction of Metro C, by year and urbanistic zone.

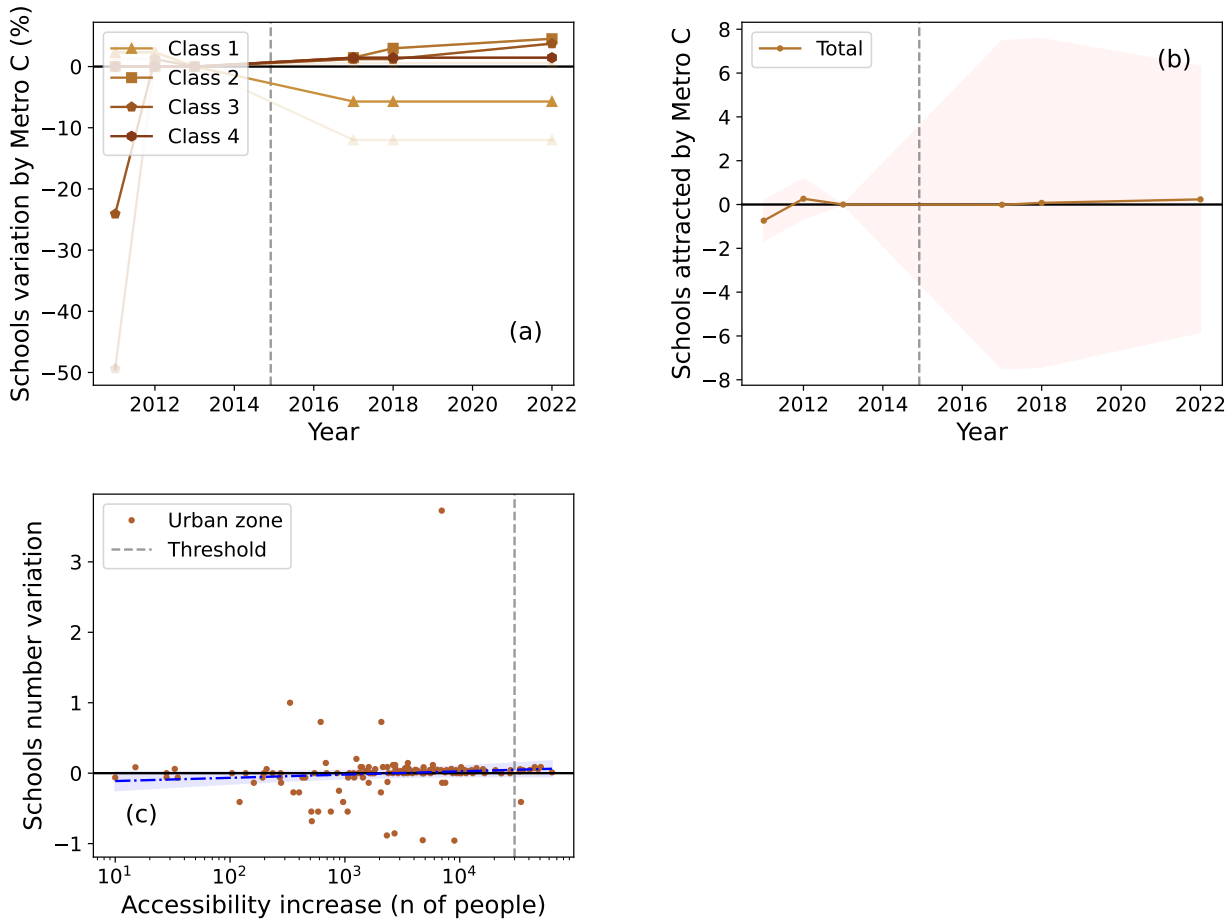

Supplementary Figure 5: **Placebo test for parallel trends assumption.** This figure replicates Fig. 2 of the main text for the placebo test for preschools. Panel (a) shows the percentage enhancement in preschools of urban zones of Rome interested by the opening of a new metro line with respect to the control group: relative difference in the average number of preschools per urban zone between zones interested by a new metro opening and zones not touched by it, as a function of the number of years after the openings. Transparent curves represent each individual urban zone. In panel (b), global absolute number of preschools affected by the Metro C opening, computed as the sum of the residuals of each urban zone affected by the new metro respect to its appropriate baseline. The shadowed region span a one standard deviation confidence interval. A null variation in preschools caused by Metro C is largely compatible with the estimation for all years under study. In panel (c), the relationship between the number variation of preschools due to the introduction of Metro C and the accessibility increase brought by it, for urban zones of Rome. Each dot represents an urban zone, the blue line is a kernel non-parametric regression, the shadowed region corresponds to a 95% confidence interval on the regression, and treated zones occupy the region of the plot on the right of the dashed gray line, while control zones are on the left. For every accessibility increase value, the kernel non-parametric regression is compatible with a null variation of preschools.

# Difference in Differences estimations of the effects of Metro C on economic activities

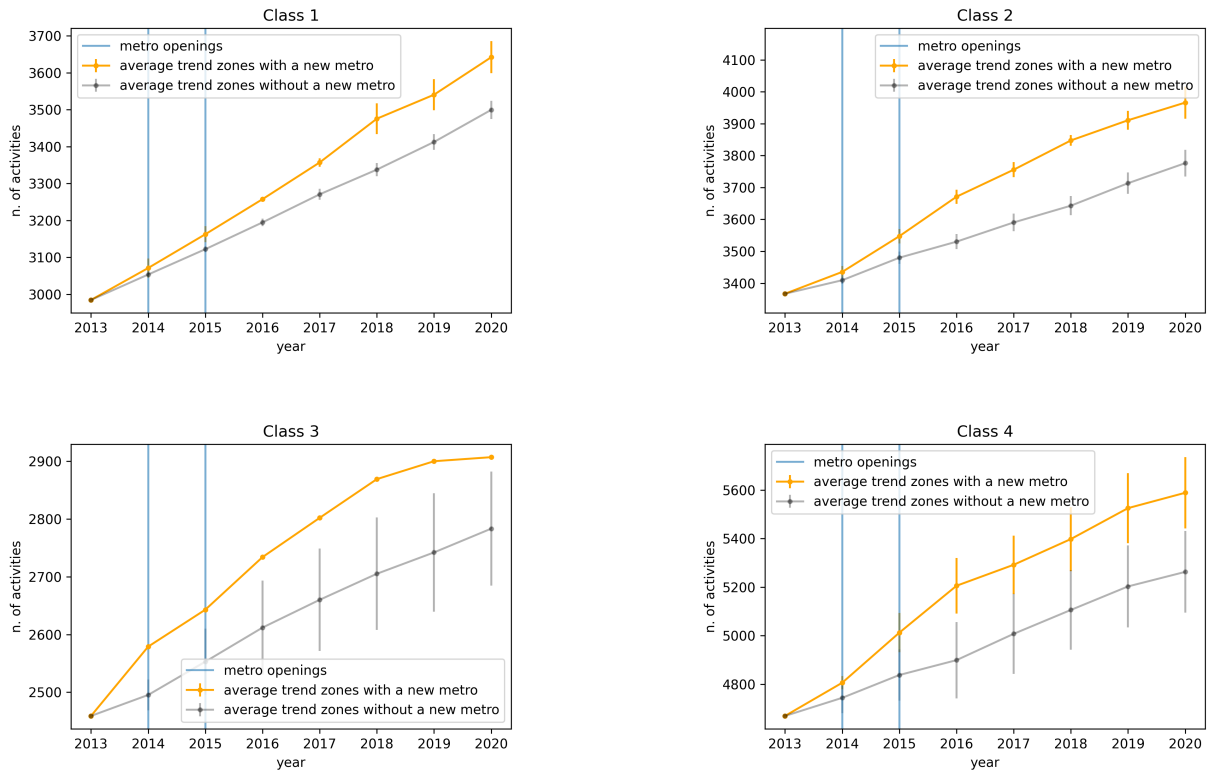

Supplementary Figure 6: Average trend (orange) compared to the corresponding baseline (grey) for the number of economic activities in urban zones affected by the opening of Metro C, grouped by economic class.

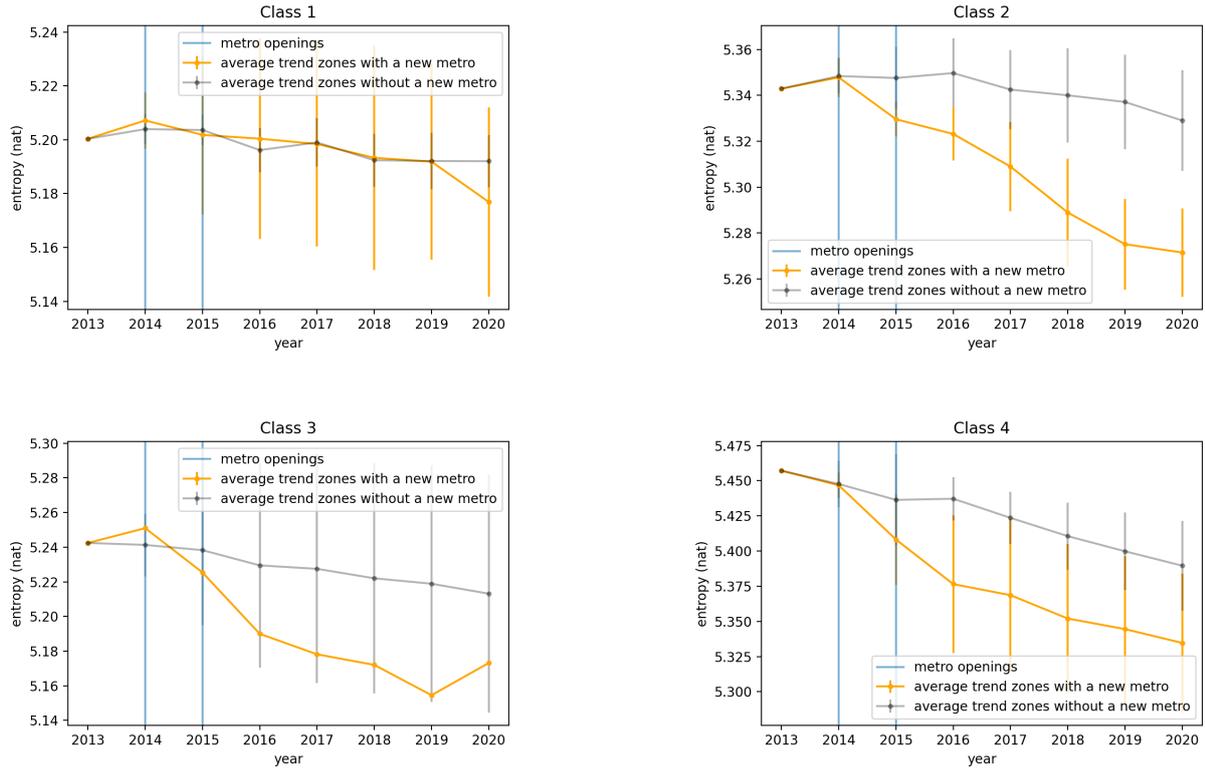

Supplementary Figure 7: Average trend (orange) compared to the corresponding baseline (grey) for the Shannon entropy $H$ in urban zones belonging to a certain economic activity density class.

|  | 2014 | 2015 | 2016 | 2017 | 2018 | 2019 | 2020 |
|---|---|---|---|---|---|---|---|
| Giardinetti-Tor Vergata | $32 \pm 43$ | $6 \pm 52$ | $34 \pm 55$ | $41 \pm 78$ | $46 \pm 111$ | $40 \pm 127$ | $47 \pm 141$ |
| Borghesiana | $4 \pm 43$ | $75 \pm 52$ | $92 \pm 55$ | $132 \pm 78$ | $230 \pm 111$ | $216 \pm 127$ | $238 \pm 141$ |
| Casilino | $22 \pm 37$ | $36 \pm 51$ | $59 \pm 56$ | $65 \pm 63$ | $60 \pm 60$ | $42 \pm 77$ | $18 \pm 113$ |
| Torre Maura | $1 \pm 37$ | $18 \pm 51$ | $110 \pm 56$ | $130 \pm 63$ | $150 \pm 60$ | $156 \pm 77$ | $158 \pm 113$ |
| Torre Angela | $56 \pm 37$ | $149 \pm 51$ | $254 \pm 56$ | $300 \pm 63$ | $403 \pm 60$ | $393 \pm 77$ | $394 \pm 113$ |
| Alessandrina | $84 \pm 27$ | $90 \pm 57$ | $122 \pm 81$ | $142 \pm 89$ | $164 \pm 97$ | $158 \pm 102$ | $124 \pm 99$ |
| Centocelle | $71 \pm 138$ | $231 \pm 266$ | $439 \pm 390$ | $437 \pm 409$ | $506 \pm 421$ | $603 \pm 445$ | $702 \pm 447$ |
| Gordiani | $24 \pm 138$ | $64 \pm 266$ | $126 \pm 390$ | $120 \pm 409$ | $113 \pm 421$ | $125 \pm 445$ | $112 \pm 447$ |
| Tuscolano Nord | $-6 \pm 138$ | $-42 \pm 266$ | $-17 \pm 390$ | $-68 \pm 409$ | $-105 \pm 421$ | $-121 \pm 445$ | $-125 \pm 447$ |
| Torpignattara | $160 \pm 138$ | $446 \pm 266$ | $678 \pm 390$ | $650 \pm 409$ | $656 \pm 421$ | $684 \pm 445$ | $616 \pm 447$ |

Supplementary Table 2: Enterprises variation due to the incroduction of metro C, for year and urbanistic zone.

## DiD estimations on microenterprises

|  | 2014 | 2015 | 2016 | 2017 | 2018 | 2019 | 2020 |
|---:|---:|---:|---:|---:|---:|---:|---:|
| Giardinetti-Tor Vergata | 31 | 9 | 43 | 48 | 51 | 48 | 51 |
| Borghesiana | 13 | 93 | 104 | 157 | 238 | 229 | 244 |
| Casilino | 16 | 27 | 53 | 63 | 58 | 45 | 21 |
| Torre Maura | -7 | 15 | 95 | 101 | 112 | 116 | 109 |
| Torre Angela | 69 | 167 | 275 | 336 | 432 | 430 | 438 |
| Alessandrina | 72 | 82 | 108 | 129 | 152 | 152 | 112 |
| Centocelle | 83 | 264 | 493 | 505 | 580 | 680 | 770 |
| Gordiani | 37 | 81 | 147 | 146 | 144 | 166 | 149 |
| Tuscolano Nord | 1 | -34 | 8 | -25 | -47 | -54 | -53 |
| Torpignattara | 170 | 466 | 722 | 705 | 708 | 750 | 674 |

Supplementary Table 3: Number of micro-enterprises created or displaced by the introduction of the metro C per year per urbanistic zone.

|  | 2014 | 2015 | 2016 | 2017 | 2018 | 2019 |
|---:|---:|---:|---:|---:|---:|---:|
| Giardinetti-Tor Vergata | 57 | 17 | 80 | 87 | 94 | 91 |
| Borghesiana | 23 | 173 | 200 | 291 | 446 | 438 |
| Casilino | 28 | 47 | 98 | 113 | 108 | 86 |
| Torre Maura | -13 | 28 | 184 | 191 | 211 | 225 |
| Torre Angela | 121 | 298 | 504 | 607 | 784 | 800 |
| Alessandrina | 123 | 143 | 195 | 231 | 277 | 285 |
| Centocelle | 142 | 460 | 898 | 912 | 1066 | 1289 |
| Gordiani | 67 | 149 | 278 | 270 | 271 | 318 |
| Tuscolano Nord | 2 | -66 | 17 | -49 | -93 | -111 |
| Torpignattara | 275 | 760 | 1238 | 1204 | 1253 | 1373 |

Supplementary Table 4: Number of workplaces in micro-enterprises created or displaced as a consequence of the introduction of the metro C per year per urbanistic zone.

|  | 2014 (M) | 2015 (M) | 2016 (M) | 2017 (M) | 2018 (M) | 2019 (M) |
|---:|---:|---:|---:|---:|---:|---:|
| Giardinetti-Tor Vergata | 1.34 | 0.04 | 0.49 | 0.54 | 0.83 | 0.36 |
| Borghesiana | 0.27 | 1.61 | 2.90 | 5.50 | 8.95 | 6.93 |
| Casilino | 0.01 | 0.14 | 0.75 | 0.73 | 0.75 | 0.01 |
| Torre Maura | -0.03 | 0.20 | 2.26 | 3.63 | 4.54 | 4.87 |
| Torre Angela | 1.93 | 3.56 | 7.12 | 13.04 | 20.67 | 14.76 |
| Alessandrina | 0.57 | 0.45 | 1.10 | 1.97 | 2.41 | 1.61 |
| Centocelle | 1.29 | 3.04 | 8.07 | 8.58 | 12.49 | 16.29 |
| Gordiani | 0.68 | 1.30 | 2.79 | 3.50 | 2.49 | 2.11 |
| Tuscolano Nord | 0.01 | -0.81 | 0.28 | -0.83 | -1.78 | -1.55 |
| Torpignattara | 1.51 | 4.09 | 6.60 | 11.42 | 12.05 | 15.23 |

Supplementary Table 5: Variation in GDP, generated by micro-enterprises, determined by the introduction of the metro C per year per urbanistic zone. All values are in millions of euros.

# Ateco codes descriptions

| ATECO code | Description |
|:---:|:---:|
| 01 | Crop and animal production, hunting and related service activities |
| 02 | Forestry and logging |
| 03 | Fishing and aquaculture |
| 05 | Mining of coal and lignite |
| 06 | Extraction of crude petroleum and natural gas |
| 07 | Mining of metal ores |
| 08 | Other mining and quarrying |
| 09 | Mining support service activities |
| 10 | Manufacture of food products |
| 11 | Manufacture of beverages |
| 12 | Manufacture of tobacco products |
| 13 | Manufacture of textiles |
| 14 | Manufacture of wearing apparel |
| 15 | Manufacture of leather and related products |
| 16 | Manufacture of wood |
| 17 | Manufacture of paper and paper products |
| 18 | Printing and reproduction of recorded media |
| 19 | Manufacture of coke and refined petroleum products |
| 20 | Manufacture of chemicals and chemical products |
| 21 | Manufacture of basic pharmaceutical products and pharmaceutical preparations |
| 22 | Manufacture of rubber and plastics products |
| 23 | Manufacture of other non-metallic mineral products |
| 24 | Manufacture of basic metals |
| 25 | Manufacture of fabricated metal products, except machinery and equipment |
| 26 | Manufacture of computer, electronic and optical products |
| 27 | Manufacture of electrical equipment |
| 28 | Manufacture of machinery and equipment n.e.c. |
| 29 | Manufacture of motor vehicles, trailers and semi-trailers |
| 30 | Manufacture of other transport equipment |
| 31 | Manufacture of furniture |
| 32 | Other manufacturing |
| 33. | Repair and installation of machinery and equipment |
| 35. | Electricity, gas, steam and air conditioning supply |
| 36. | Water collection, treatment and supply |
| 37. | Sewerage |
| 38. | Waste collection, treatment and disposal activities; materials recovery |
| 39. | Remediation activities and other waste management services |
| 41 | Construction of buildings |
| 42. | Civil engineering |
| 43 | Specialized construction activities |
| 45 | Wholesale and retail trade and repair of motor vehicles and motorcycles |
| 46 | Wholesale trade, except of motor vehicles and motorcycles |
| 47 | Retail trade, except of motor vehicles and motorcycles |
| 49. | Land transport and transport via pipelines |
| 50. | Water transport |
| 51. | Air transport |
| 52. | Warehousing and support activities for transportation |
| 53. | Postal and courier activities |
| 55 | Accommodation |
| 56 | Food and beverage service activities |
| 58. | Publishing activities |
| 59. | Motion picture, sound recording and music publishing activities |
| 60. | Programming and broadcasting activities |
| 61. | Telecommunications |
| 62. | Computer programming, consultancy and related activities |
| 63. | Information service activities |
| 64. | Financial service activities, except insurance and pension funding |
| 65. | Insurance, reinsurance and pension funding, except compulsory social security |
| 66. | Activities auxiliary to financial services and insurance activities |
| 68 | Real estate activities |
| 69. | Legal and accounting activities |
| 70. | Activities of head offices; management consultancy activities |
| 71. | Architectural and engineering activities; technical testing and analysis |
| 72. | Scientific research and development |
| 73 | Advertising and market research |
| 74. | Other professional, scientific and technical activities |
| 75. | Veterinary activities |
| 77 | Rental and leasing activities |
| 78 | Employment activities |
| 79 | Travel agency, tour operator, reservation service and related activities |
| 80 | Security and investigation activities |
| 81 | Services to buildings and landscape activities |
| 82 | Office administrative, office support and other business support activities |
| 84. | Public administration and defence; compulsory social security |
| 85. | Education |
| 86. | Human health activities |
| 87. | Residential care activities |
| 88. | Social work activities without accommodation |
| 90 | Creative, arts and entertainment activities |
| 91 | Libraries, archives, museums and other cultural activities |
| 92 | Gambling and betting activities |
| 93 | Sports activities and amusement and recreation activities |
| 94. | Activities of membership organisations |
| 95 | Repair of computers and personal and household goods |
| 96 | Other personal service activities |
| 97. | Activities of households as employers of domestic personnel |
| 98 | Undifferentiated goods- and services-producing activities of private households for own use |
| 99 | Activities of extraterritorial organizations and bodies |

# Estimation of economic sectors favoured and disadvantaged by the introduction of Metro C

For each economic class, we computed the average difference $d$ in the number of enterprises per ATECO code between 6 years after the opening of the metro C and the year previous to the opening, separately for treated $d_T$ and control zones $d_C$. In panels (a) of Figures from 8 to 11, the average over control zones is shown in orange, and the average over control zones is shown in gray. Panels (b) show the relative difference $\Delta$ between the orange curve and the gray one

$$\Delta = \frac{d_T - d_C}{d_C}\,.$$

This quantity provides an estimation of the effect of Metro C on economic activities belonging to each ateco code, six years after its opening.

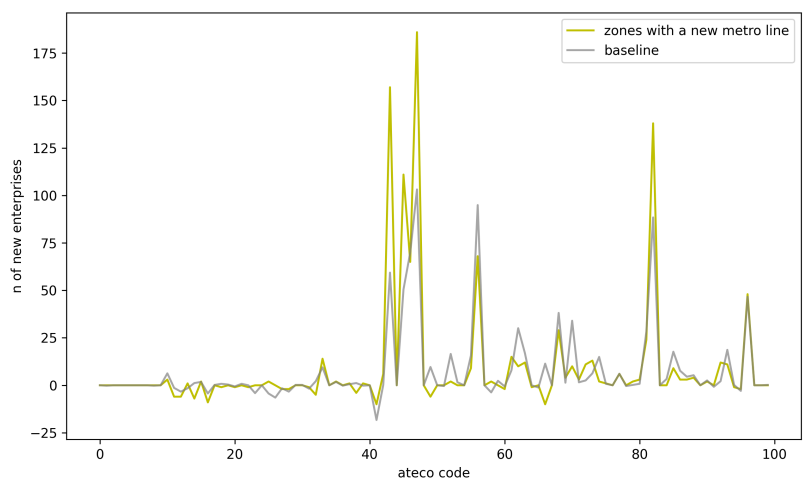

(a) Average difference in the number of enterprises per ATECO code between 6 years after the opening of Metro C and the year prior to the opening, averaged across class 1 urbanistic zones.

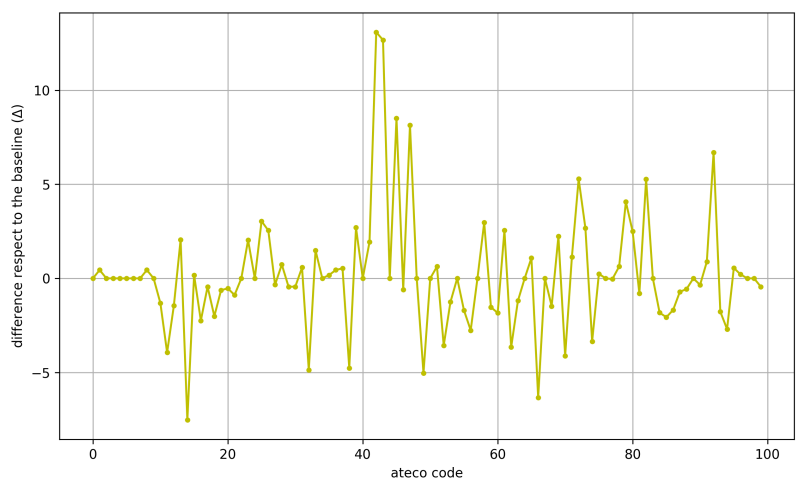

(b) Variation with respect to the baseline over 6 years in urbanistic zones affected by the opening of a new metro line.

Supplementary Figure 8: Effect of Metro C on the distribution of economic activities by sector in class 1 urbanistic zones.

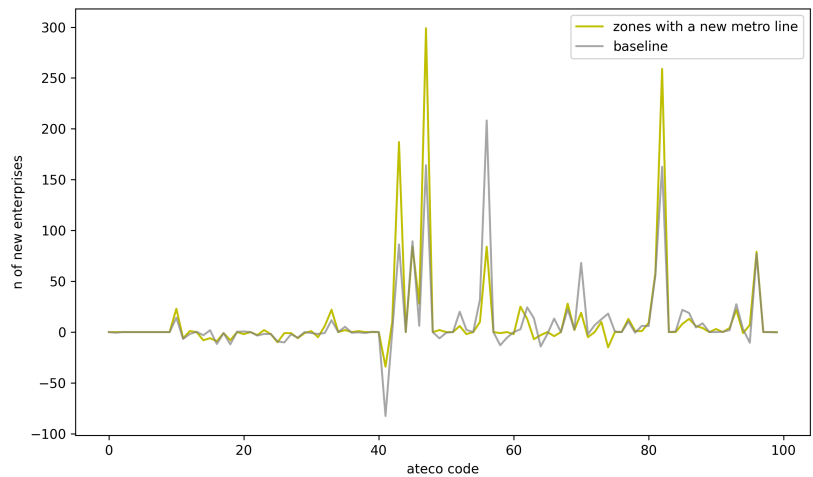

(a) Average difference in the number of enterprises per ATECO code between 6 years after the opening of Metro C and the year prior to the opening, averaged across class 2 urbanistic zones.

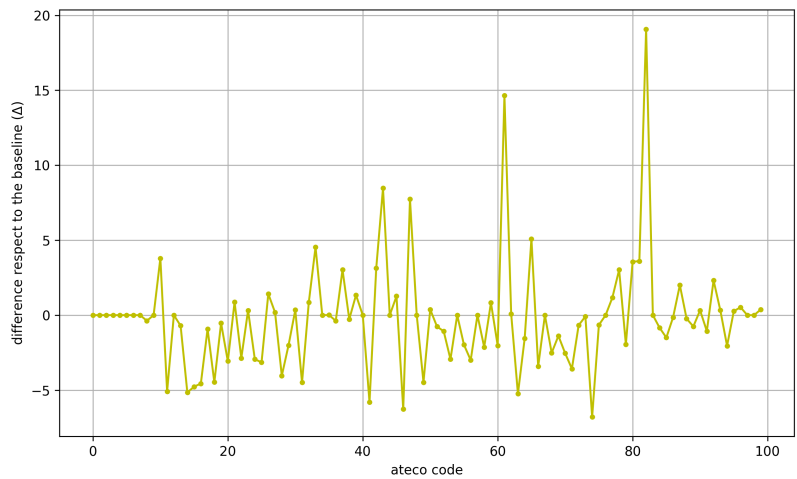

(b) Variation with respect to the baseline over 6 years in urbanistic zones affected by the opening of a new metro line.

Supplementary Figure 9: Effect of Metro C on the distribution of economic activities by sector in class 2 urbanistic zones.

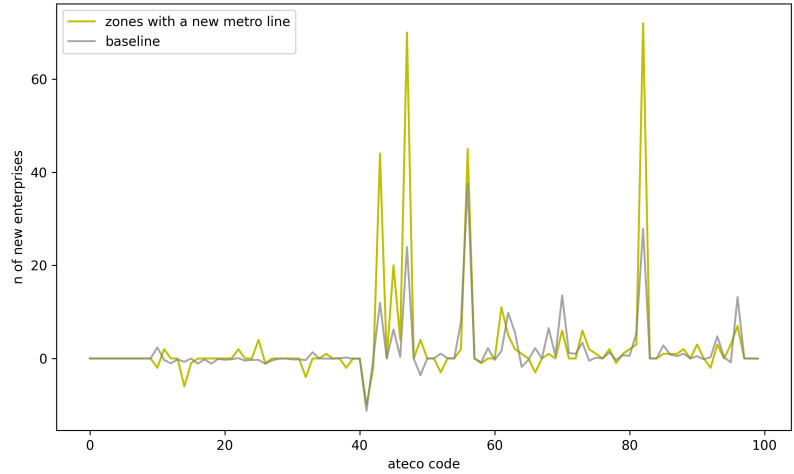

(a) Average difference in the number of enterprises per ATECO code between 6 years after the opening of Metro C and the year prior to the opening, averaged across class 3 urbanistic zones.

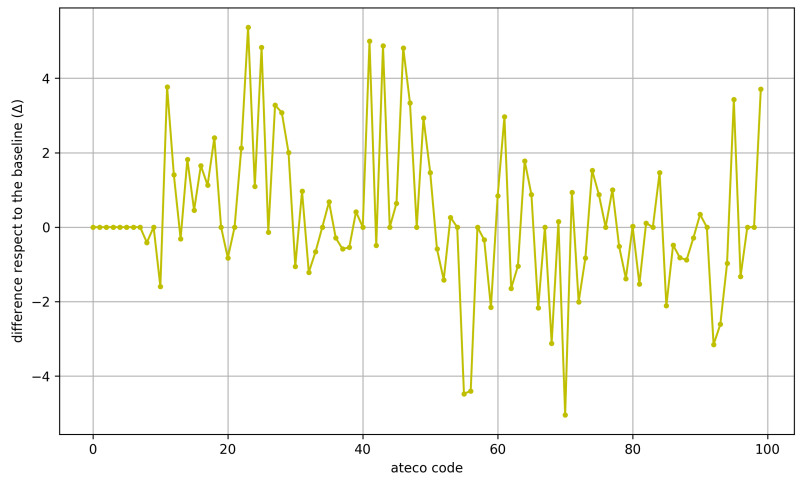

(b) Variation with respect to the baseline over 6 years in urbanistic zones affected by the opening of a new metro line.

Supplementary Figure 10: Effect of Metro C on the distribution of economic activities by sector in class 3 urbanistic zones.

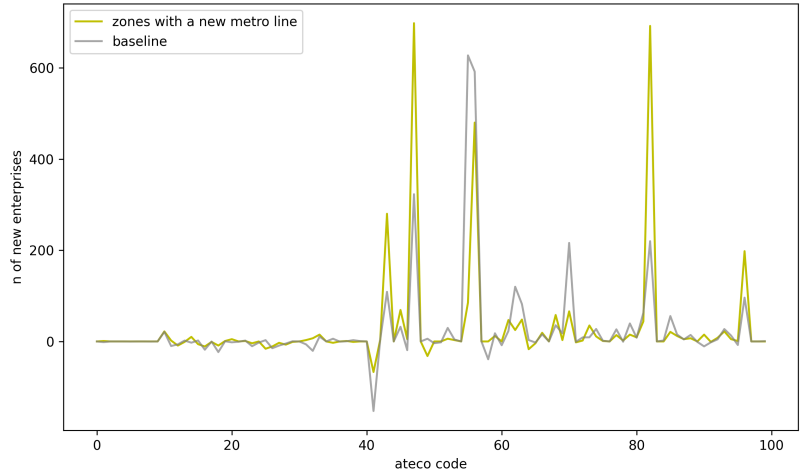

(a) Average difference in the number of enterprises per ATECO code between 6 years after the opening of Metro C and the year prior to the opening, averaged across class 4 urbanistic zones.

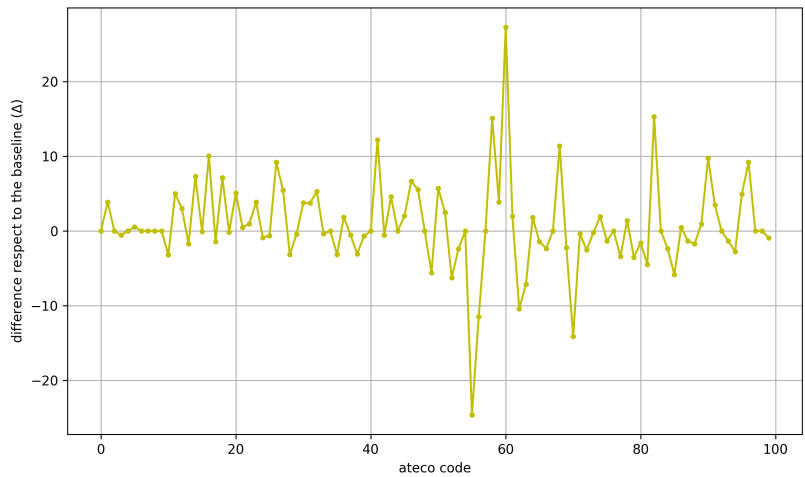

(b) Variation with respect to the baseline over 6 years in urbanistic zones affected by the opening of a new metro line.

Supplementary Figure 11: Effect of Metro C on the distribution of economic activities by sector in class 4 urbanistic zones.